\documentclass{aastex}
\shorttitle{Variable Stars in 47 Tucanae}
\shortauthors{Weldrake et al.}

\begin{document}
\title{A Comprehensive Catalogue of Variable Stars in the field of 47 Tucanae.}

\author{David T F Weldrake} 
\affil{Research School of Astronomy and Astrophysics, Mount Stromlo Observatory, Cotter Road, Weston Creek, ACT 2611, Australia}
\email{dtfw@mso.anu.edu.au}

\author{Penny D Sackett}
\affil{Research School of Astronomy and Astrophysics, Mount Stromlo Observatory, Cotter Road, Weston Creek, ACT 2611, Australia}
\email{psackett@mso.anu.edu.au}

\author{Terry J Bridges}
\affil{Anglo-Australian Observatory, P.O. Box 296, Epping. NSW, 1710, Australia}
\affil{Current Address: Department of Physics, Queen's University, Kingston, Ontario. 
\linebreak K7L 3N6, Canada}
\email{tjb@astro.queensu.ca}
\and

\author{Kenneth C Freeman} 
\affil{Research School of Astronomy and Astrophysics, Mount Stromlo Observatory, Cotter Road, Weston Creek, ACT 2611, Australia}
\email{kcf@mso.anu.edu.au}

\begin{abstract}
We present the results of a comprehensive search for stellar variability in the globular cluster 47 Tucanae. Using the Mount Stromlo 40-inch (1m) telescope at Siding Spring Observatory and a combined V+R filter, we have detected 100 variable stars across a 52$\times$52$'$ field centered on the cluster. The main aim of this project is to search for transiting 'Hot Jupiter' planets in this cluster, the results of which shall be discussed in a separate paper. Here we present the V+R lightcurves and preliminary investigations of the detected variable stars, which comprise 28 Eclipsing Binaries (21 contact binaries and 7 detached systems), 45 RR Lyrae stars (41 of which belong to the Small Magellanic Cloud and four seemingly to the Galactic Halo), and 20 K-giant Long Period Variables (LPVs). We also detected four $\delta$ Scuti stars, one TypeI Cepheid, and one TypeII Cepheid. One variable appears to be a possible dust-enshrouded SMC star with a short period pulsation. Of these 100 variables, 69 are new discoveries. Our eclipsing binary sample indicates a clear radial segregation in period, and includes two binaries that are seemingly orbited by low-luminosity stellar companions. One RR Lyrae star shows a Blahzko effect with remarkable regularity.

Those variables previously known are cross-identified with Kaluzny and coworkers. In agreement with previous studies, this work shows that the relative frequency of detectable variable stars (particularly contact binaries) in 47 Tuc is very low compared to other studied regions. A distance modulus of 18.93$\pm$0.24 for the Small Magellanic Cloud and 13.14$\pm$0.25 for 47 Tucanae has been estimated from our sample, and are in agreement with values previously published.

The total database presented here contains V and I photometry for 43,067 47 Tuc field stars (13.0$\leqslant$V$\leqslant$21.0), along with 33-night V+R lightcurves and astrometry for 109,866 stars (14.5$\leqslant$V$\leqslant$22.5).  
\end{abstract}


\section{Introduction}
Variable stars in globular clusters (particularly binaries) play an important role in understanding cluster dynamical evolution. Despite this, such clusters have seldom been the targets of detailed study for the presence of variables, mainly due to the difficulty in obtaining accurate photometry from the ground of faint stars in very crowded fields. Due to technological advances in recent years, however, the number of clusters studied and the list of variable stars discovered has increased dramatically. As a general overview of the field, those major clusters investigated recently include Omega Centauri \citep{Kal96,Kal97,Hag03}, M5 \citep{YR96}, M71 \citep{YM94}, M4 \citep{KTK97}, NGC6397 \citep{K97}, M22 \citep{Piet03}, NGC6946 \citep{Guld03}, M69 \citep{Gregorsok03}, M15 \citep{Zhel03}, M13 \citep{Kopacki03} and M3 \citep{Strader02}. \citet{YC96} discovered six spectroscopic binaries in NGC5053. \citet{Clement01} and \citet{H96} review recent and more historical studies into globular variables.

A few previous surveys have searched for variable stars in 47 Tuc. \citet{Hogg73} discovered two variables. \citet{Edmonds96a,Edmonds96b} used a HST dataset to detect 75 variable stars, including Eclipsing Binaries and variability among K-giants. \citet{Alb01} uncovered 107 variable stars, the largest number to date, and derived an overall binary frequency of 14$\%\pm$4$\%$ using the same HST dataset as \citet{Gil2000} to search for Hot Jupiter planets in the cluster. \citet{Kal98} performed a wider field survey on the cluster, uncovering 42 variables. 

We present a new extensive variable star catalogue, a natural byproduct of a photometric project to search for transiting 'Hot Jupiter' planets in 47 Tuc. The results of the planetary search will be the subject of a separate paper. The total catalogue comprises 28 Eclipsing Binary systems, 20 long period variables and 41 Small Magellanic Cloud (SMC) RR Lyraes, four Halo RR Lyraes, two Cepheids and four apparent $\delta$ Scuti stars, and one anomalous short period Small Magellanic Cloud (SMC) star. Discrimination between cluster/SMC memberships is achieved using the location of the variable on the cluster Colour Magnitude Diagram. Our large number of new discoveries is due primarily to the very large field of view (52$\times$52$'$) of our survey and also its photometric depth. In this paper, we present the phase-wrapped V+R lightcurves, preliminary investigations into the detected variable stars, and a description of our photometric, astrometric and lightcurve database.

Our survey covers a larger area than any previous search, and extends to deeper photometry than presented by \citet{Kal98}. We recover 31 of Kaluzny's stars and discover a further 69 variables. The unrecovered variables either lie between CCDs, or within regions of no data caused by telescope offsets. The cluster core cannot be easily imaged by ground-based telescopes due to the extreme crowding. On our 300s exposures (V+R), the inner 6$'$ of 47 Tuc is saturated. The cluster field is located at a high galactic latitude (l=305.9 deg, b=$-$44.9 deg), providing low foreground contamination by the Milky Way and low reddening. The field is significantly contaminated by background stars from the Small Magellanic Cloud. Our field of view extends to $\sim$60$\%$ of the 47 Tuc tidal radius. Study of the SMC RRLyrae stars, as standard candles, presents an opportunity to investigate the distance to the SMC from a location some seven degrees NW of the centre.

Contact eclipsing binaries (EcB) are very useful as distance indicators. Observing any double line spectroscopic binaries offers an opportunity to measure directly the primary stellar parameters such as mass, luminosity, radius and hence distance. \citet{Ruc93} presents a method of distance determination for contact binaries by including the period, unreddened V-I colour and system metallicity. We present the results of an application of Rucinski's calibration for our binary sample with periods less than 1 day.

Section 2 describes the observations and data reduction, along with a description of the method used to obtain the lightcurves. Here we also discuss the astrometry and variable detection methodology, along with notes on cluster membership of the variables. Our survey completeness and the quality of the photometry is described. Section 3 deals with the cluster Colour Magnitude Diagram (CMD), the corresponding photometric database, and describes the photometric calibrations. Section 4 contains a discussion and preliminary analysis of the different types of variables in the catalogue and presents their lightcurves. We summarise and present conclusions in Section 5. Finder charts of these stars are provided on a webpage link to allow easy identification on 47 Tuc wide field images.

\section{Observations and Data Reduction}
This project was undertaken using the Australian National University (ANU) 40-inch (1m) telescope at Siding Spring Observatory. We used the Wide Field Imager (WFI) which consists of a 4$\times$2 array of 2048$\times$4096 pixel back-illuminated CCDs. The CCDs are arranged to give a total format of 8K$\times$8K pixels. The detector scale is 0.38$"$/pix at the 1m telescope Cassegrain focus, with a field of view of 52 arcmin on a side. This very large field allows us to image much of 47 Tuc, approaching 60$\%$ of the cluster tidal radius of 45.9 arcminutes \citep{Leon2000} in one exposure. The field layout is presented in Fig.\ref{astromplot}. Our main aim of detecting planetary transits requires an effective SN ratio of 200 or more at V=18 for a 3$\sigma$ detection and, to maximise the in-transit sampling, exposures no longer than 300s were employed. Our broadband filter covers the combined wavelength range of Cousins V and R, giving a significant increase in signal-to-noise while maintaining the image degradation due to atmospheric dispersion at an undetectable level. From previous experience with this filter, for a star of V=18.5 in 2$''$ seeing, the photon noise S/N decreases from 220 at 7-day moon to 165 at bright moon.  

The globular cluster 47 Tucanae was observed for a total of 33 nights from 2002 August 22 to 2002 September 24 with a field centre of RA=00h24m05.2s DEC=$-$72$^{\circ}$04$'$51.0$''$. Approximately 80$\%$ of the observing time was useful for the main planetary transit project, with mean seeing of 2.2 arcsec. The temporal coverage of the cluster was maximised as much as possible; and averaged an image every 6 minutes for around 10 hours per night. Each image was checked for quality independently after readout. Unsuitable images caused by satellite trails, bad seeing periods, etc, were discarded. 

In total we have 1220 images of the same field centered on 47 Tuc, which have been used to produce time-series lightcurves for 109,866 unsaturated stars, with apparent magnitude 14.5$\leqslant$V$\leqslant$22.5. This covers a large range of stellar mass and type, encompassing most of the red giant branch (RGB) the subgiant branch, the cluster turn-off, and the cluster main sequence down to a magnitude of V=22.5. This dataset therefore covers a large range of variable types.

Initial reduction of the raw images was undertaken using the MSCRED package of IRAF \footnote{IRAF is distributed by National Optical Observatories, which is operated by teh Association of Universities for Research in Astronomy, Inc., under cooperative agreement with the National Science Foundation}. A significant number of calibration images were obtained over the 33-night run, and allowed for correction of time-dependent variations, eg., differences in flat fields from one observing period to the next, across the WFI array. Initial reductions included region trimming and overscan correction, bias correction, flat-fielding and dark current subtraction. A dark-sky flat was obtained, but was not used because it degraded the large scale image quality. After checking  the final reduced images for flatness (using IMSTAT) and pixel-to-pixel variations, the resultant 1220 object images were then ready for the main photometric pipeline and analysis. 

\subsection{Photometry}
The primary photometric method to generate lightcurves was carried out using a Difference Image Analysis (DIA) method originally described as a optimal PSF matching algorithm by \citet{AL98}, and modified by \citet{Woz2000}. Detailed discussion of each program making up the pipeline can be found in Wozniak's paper. We summarize the method here. DIA consists of the following steps: Firstly, all frames are transformed onto a common coordinate system. Then a template image is produced by combining the best quality images with small offsets into a master image. The Point Spread Function (PSF) of the stars on each image is determined using a combination of two Gaussians, one for the core and one a factor of 1.83 wider for the wings, each multiplied by a third order polynomial. The best PSF-matching kernel is found and each image is then subtracted from the template (reference) frame. The coordinates of all stars on the template images are found, and finally the profile photometry is extracted from those star positions.

Central to this method is the assumption that the pixels containing PSF information in the analysis image vary only slightly due to seeing variations. $\it{A priori}$ knowledge of the PSF and the sky background is not required, and the method works better as the crowding increases, as in denser fields a larger number of pixels contain information about any PSF differences. Heavy blending can cause trouble, however, as the pipeline photometry does not model surrounding stars and thus faint and/or heavily blended objects can be contaminated by neighbouring stars. For this reason, a crowding flag is introduced: a flag of 1 indicates that a pixel brighter than 0.15 $\times$ distance in pixels $\times$ pixel flux lies with in the immediate four pixel neighbourhood. For the production of the PSF, a star is rejected if another local maximum at least 2$\sigma$ above the background level is present. 

For ease of data handling, each of the eight WFI CCDs were cut into 32 subsections of 512$\times$512 pixels each. Once the images were reduced, the 45 exposures with the best mean seeing, measured at 1.1$''$ were then coadded to create a template frame, a very high signal-to-noise version of each subsection. This allowed initial flux measurements of the stars to be made, which defines the zero point of the output lightcurves. The coordinates of all the stars detected by DAOFIND on these 512$\times$512 template images were fed into the main photometric pipeline. Those stars within 22 pixels of the subframe edge were discarded; due to telescope pointing offsets those stars do not have complete temporal coverage. Due to these offsets, $\sim$21,000 stars ($\sim$15$\%$ of total) could not be produced for the final lightcurve database.  

After the subtraction process is complete, the derived centroid of any variable object is unbiased by surrounding unvarying stars. The DIA method measures the flux differences between the frames, rather than the total flux. The lightcurve of the variable object can be converted to total flux units using the stellar flux from the template. 

As originally implemented, the method automatically detects objects that are classified as variable from frame to frame using the following conditions: At least three consecutive points depart by at least 3$\sigma$ from the baseline in the same sense, and there are at least 10 points departing by 4$\sigma$ from the baseline in either sense. The method then produces a lightcurve at the pixel coordinates of the detected varying object. As the main aim of the project was to find the periodic $\sim$1-2$\%$ drop in brightness caused by a planetary transit, which is practically undetectable by this automatic process, we altered the method slightly. The program was forced to produce a lightcurve for every visible star on our best-seeing image, which yielded the largest number of stars of any of our dataset images. Using DAOFIND within IRAF, the pixel locations of all stars were determined. The main photometric pipeline within DIA was then instructed to produce a lightcurve for each of these locations. The variable stars presented in this paper were detected using this modified method. This implementation of the method gave us a total 1200-point lightcurve database across the 47 Tuc field for 109,866 unsaturated stars. This is a smaller number than that of our total astrometric database, due to telescope offsets and the gaps between images.

The DIA method produces lightcurves that are not on a standard magnitude system. The lightcurves are produced with linear flux units, from which a constant reference flux taken from the template images has been subtracted. To convert this unit into a magnitude scale, the flux of each star on the reference image must be determined. An estimate of the flux was made by running DAOPHOT on the reference template images and measuring the flux (in counts) for each star. The change in light output in magnitudes can then be calculated via: 

\begin{center}
$\Delta m_i = -2.5 \log [(N_i+ N_{\rm ref,\it{i}}) / N_{\rm ref,\it{i}}]$
\end{center}

\noindent where $N_{\rm ref,\it{i}}$ is the flux of star $\it{i}$ on the template image, and $N_i$ is the difference flux on a given frame in the time series. We then compared our range of $\Delta m_i$ for those variables found by \citet{Kal98} for verification. These $\Delta m_i$ are thus used as the magnitude total variability level of our detected variables. Fig.\ref{ampmags} shows the amplitude comparison between the two datasets for those variables which can be crossidentified, and shows a scatter around the zero-point with a mean of -0.0016 and a standard deviation of 0.13 magnitudes. The slope of the LPV variables towards fainter magnitudes is attributed to the different passbands of \citet{Kal98} (V) and ourselves (V+R). Those EcB which are not close to the zeropoint are members of the binary main sequence, and hence are more likely to be composed of different colour components. The RR Lyrae scatter is attributed to the relative faintness of this sample.

Our photometric errors are presented in Fig.\ref{rmstot}, which shows the V+R rms photometric uncertainties of our lightcurves as a function of V magnitude. As we have a significant change in crowding between the inner four CCDs and the outer four (see Fig.\ref{astromplot}), two rms curves are presented. It can be seen that the outer CCDs produce comparable lightcurve quality for stars up to 0.5mag fainter for any given rms level compared to the inner CCDs. We therefore have sufficient photometric precision to detect a 1.5-2.5$\%$ dip (typical of the general range of a 'Hot Jupiter' planet \citep{Char2000} in the lightcurves of V=18.5 stars in a crowded inner CCD and V=19.0 in an uncrowded outer CCD. This limit allows us to probe the brightest 1.5-2 magnitudes of the cluster main sequence for any orbiting giant planets; these results will be presented in a subsequent paper. Our rms precision also allows us to detect variable stars to V$\sim$21 with amplitudes of (0.14,0.20) magnitudes for the two crowding levels, giving us an excellent chance to detect previously unknown variable stars. Fig.\ref{starcount} shows the magnitude depth of our photometry for all eight CCDs, with recovered stars counted in 0.25 magnitude bins. Our photometry is limited to the range 13.0$\leqslant$V$\leqslant$22.0, with different chips having slightly different sensitivities. The apparent peaks in the stellar distributions are due to incomplete sampling at faint magnitudes. The gradual decrease in star numbers towards CCD8 can be clearly seen, indicative of increasing distance from the Small Magellanic Cloud (SMC), and hence decreasing background contamination.

\subsection{Astrometry}
Astrometry was obtained for all 143,814 stars detected in our best seeing image, across the eight CCDs of the 47 Tuc field allowing a determination of the position of all stars in both our photometric and lightcurve database. We used a program to search the USNO CCD Astrograph Catalogue (UCAC1) for astrometric standard stars within the field (B.P.Schmidt, 2003 private communication). Several hundred such stars were cross-identified, allowing for an accurate determination of the astrometric solution of our star lists. The rms residual of the astrometry was $\sim$0.15 arcsecs. Our astrometry is presented in Fig.\ref{astromplot}, with the CCD number overplotted. The cluster core and our extensive coverage of the cluster is apparent.

\section{Colour Magnitude Diagram}
The V,V-I Colour Magnitude Diagram (CMD) used in the production of the variable colour data is presented in Fig.\ref{47TucCMD_total}. These data were originally taken at the MSSSO 40$''$ telescope by Ken Freeman and Michelle Doherty to allow for a placement of any candidate transiting system on the V,V-I system. The data cover the same 52$\times$52$'$ FOV as the lightcurve dataset presented in this paper, and totalling 43,067 stars. The magnitude range is 13.0$\leqslant$V$\leqslant$21.0 in V and I. The CMD was calibrated against that presented by \citet{Kal98}. The authors warn in that paper of systematic errors caused by non-linearity of the OGLE CCD chip. For faint stars these errors are likely to be more significant. 
The OGLE dataset as available on their website (calibrated) was overplotted on top of our uncalibrated CMD data. Our data was then shifted until the two datasets overlay eachother. A histogram of the distribution of the stellar magnitudes was produced, and our data was shifted further until a more accurate match was found by comparing the V magnitude of the Horizontal Branch.  This was repeated for V-I calibration. This simple calibration method produced photometry accurate to $\leqslant$0.03 mag, across the full magnitude range. Fig.\ref{47TucCMD_total} shows our CMD. Variable stars in the V+R dataset were cross-identified by comparing the template image containing the variable and the corresponding CCD image of the CMD dataset. The V and V-I information in Tables 3,4 and 5 are associated with uncertainties of 0.03 and 0.04 respectively, and are the non-random errors in the zero-point determination and incorporate the errors in the OGLE calibration. 

\subsection{Method for variable detection.}
A Lomb-Scargle Periodogram (LSP) method was chosen to search for periodic variables in the final lightcurve database \citep{Brett01}. This method was produced specifically to overcome the problem of unevenly spaced data, which in our case is caused by daylight and cloudy nights. A Fourier power spectrum is produced, with the same statistical properties as a standard power spectrum. If a periodicity (P) is detected in the data, a spike is produced in the spectrum at a frequency of 2$\pi$/P. As implemented, any spike above 2$\times$rms of the power spectrum is identified as a candidate. Using this condition, 29,314 lightcurves which contained such an apparent periodicity ($\sim$23$\%$ total database, $\sim$300 times the final variable number) were found, which were then examined in detail. It was noticed that the vast majority of this list was composed of common systematic effects inherent in the data, lightcurves from stars close to saturation, stars very close to our magnitude limits and the $>$2$\sigma$ variables. All those candidates which exhibited a clear periodicity by eye were catalogued, leaving a sample of 100 for the variable list. Fig.\ref{detlim} shows our detection limits. The log of the total amplitude (in magnitudes) of the detected variables have been plotted against V. It is clear that there is a sharp cutoff to the detections, marked with a dotted line. Any amplitudes which are $>$3$\%$ are detectable to V=17.0, and only those with an amplitude $>$1 mag can be found for V$>$22. Hence those variables which lie underneath this limit are missed in our dataset.

By phase-wrapping the candidate lightcurves at the detected period, the nature of the periodicity is seen. The LSP-derived period is very close $\sim$0.03d to the chosen 'true' periods presented for our variables in Table1. Once phase-wrapped, the period was altered (at the third significant figure) until a minimum in the scatter of the points was seen; this value was recorded as the period of the variable. Our cross-identified variables had their derived periods compared to those in \citep{Kal98}, and are in excellent agreement.

\section{Preliminary Variable Analysis}
\subsection{Eclipsing Binaries}
Phase-wrapped lightcurves of all 100 variable stars detected in our dataset are presented in Figs.\ref{VarPlot1}-\ref{VarPlot10}. We arrange the lightcurves by type, EcB (Figs.\ref{VarPlot1}-\ref{VarPlot3}), RR Lyraes (Figs.\ref{VarPlot4}-\ref{VarPlot7}), long period variables (LPVs, Figs.\ref{VarPlot8}-\ref{VarPlot9}) and miscellaneous variables (Fig.\ref{VarPlot10}). For the EcB, it is clear that examples of contact (ie V6), semi-contact (ie V7) and detached configuration (ie V69) binaries are present, classified due to the appearance (or lack thereof) of features in the lightcurves indicating tidal distortion of the stellar components, the period, and the length of time between eclipses.  The EcB sample consists mostly of W Ursae Majoris-type lightcurves, with short periods. Most of these stars seem to be candidate blue stragglers with cluster membership (see Fig.\ref{linecmd}). The lightcurves of some of these stars (ie, V6, V7, V12) display secondary variations outside the main  variability, which could be indicative of star-spots, flares or Roche-Lobe overflow between the components (E.C Sung, 2003 private communication). One of the main explanations for the phenomenon of Blue Straggler Stars (BSS) is that the two binary components are tranferring mass, and hence remain on the main sequence when they otherwise would have evolved off. \citet{L96} showed that many BSS are members of short-period binaries, and as such are likely to be transferring material from one to the other as one component evolves. Hence it is not surprising that some of our sample of Blue Stragglers (ie V6) seem to show evidence of such mass transfer activity in their lightcurves.

A number of EcB appear to lie on the cluster Binary Main Sequence (BMS) and as such all are very likely members of 47 Tuc. Detached systems are also seen in the sample (V39, V41, V69, V78, V84, V89 and V93). Of these, four do not show detectable secondary eclipses to our rms level (V69, V78, V89, V93), and thus could conceivably be orbited by low luminosity companions, most likely M-dwarfs. Our data for V78 and V93 is very limited, with only one eclipse visible across our temporal range. Our period estimates for these two variables are therefore not well determined. We have estimated a period for these two stars which would hold if all other eclipses occured during cloud or daylight. Further observations are required to derive an accurate period. If these stars are indeed orbited by M-Dwarfs, these variables would be important to determine the survivability and long-term stability of such low-mass companions inside globular clusters \citep{AL03}. We present phase wrapped lightcurves of these two 'special cases' in Fig.\ref{Mdwarf}, with the eclipses plotted more clearly than in the general lightcurve database figures. V78 also shows an apparent variation at 1.015d, as well as the longer period for which only one eclipse is seen. This is shown in Fig.\ref{VarPlot2} (two plots for V78), and as the period is so close to 1 day, it is almost certainly due to terrestrial effects. It is included for completeness. The lightcurve data suggest that the companion sizes are approximately 0.25-0.3 Solar radius, assuming a mid-to late K-type primary at the distance of 47 Tuc. The rest of our EcB sample are detached binaries on the cluster BMS.

The apparent frequency of the occurence of detectable 47 Tuc contact binaries in the field is 21/124073 $\thickapprox$ 1.7$\pm$0.4$\times$10$^{-4}$, which is slightly higher, but consistent with, the estimate of 1$\times$10$^{-4}$ presented in \citet{Kal98}, and is more than an order of magnitude lower than that observed in the core of 47Tuc \citep{Alb01}, and in fields containing Galactic open clusters \citep{KR93} and OGLE fields located in Baade's Window, close to the Galactic Centre \citep{R97}. The reasons for this difference are unclear, although mass segregation may play an important role in determining the radial distribution of such stars. For comparison, some other globular clusters studied with variability surveys give eclipsing binary frequencies of 7.5$\times$10$^{-4}$ for M71 \citep{YM94} and 1.4$\times$10$^{-3}$ for M5 \citep{YR96}.  

Our dataset shows an apparent segregation in binary period, with shorter period binaries located closer (in projection) to the core of 47 Tuc. This can be seen clearly in Fig.\ref{EcBdist} where we have plotted the number of contact and detached EcB as a function of radius from the core in bins of 6 arcminutes. A (normalised) histogram of the total star population has been overplotted for comparison. The two EcB distributions are clearly different, with a significance of around 4$\sigma$, as determined by a Kolmogorov Smirnov test. Contact systems are preferentially located closer to the core (in projection) than detached systems of the same magnitude. This segregation is not a classical indication of mass segregation, for which we would expect both detached and contact binary systems to be preferentially located closer to the core, but may be an indication of dynamical effects, with binaries closer to the core having lost much of their orbital energy. The binaries further out retain their energy, and hence remain in longer period orbits.  Due to dynamical effects, it is expected that most of the cluster contact binaries would be located in the core \citep{H92}, although some would have been ejected from the core by star-star encounters. A theoretical King Profile has also been overplotted to show the total stellar distribution, without the crowding our data suffers towards the core.
The inner 6$'$ is not sampled in our dataset; the dropoff on binary number at the inner 6 arcminutes is entirely due to this effect.   

Fig.\ref{EcBPcol} shows the sample of our contact EcB (periods $\leqslant$0.5d) indicating a clear relationship between Period and V-I colour. This period-colour relationship was first noted by \citet{Eggen67}, and further investigated by \citet{Rub2001}. The explanation as offered by these authors is that this relation is indicative of the differing system masses, the higher mass systems are bluer and have longer orbital periods than the lower mass systems. As one moves further down the cluster CMD (in our case), the mass of the system in total decreases, and the overall system colour gets redder. Since massive stars are physically larger than low-mass stars, those binaries which have high mass components must have a greater orbital separation to accommodate this larger stellar volume. Therefore the lower the likely masses of the components the redder they appear, and the shorter the orbital period will be. System metallicity has a bearing on the positions of the contact binaries in the period colour relationship. \citet{YM94} showed that the reduced line blanketing of metal-poor stellar atmospheres accounted for some of the differences between a globular cluster period colour relationship and that of field binaries. The lower metallicity also makes stellar radii smaller, affecting the resultant orbital period for such systems. 

\subsection{Notes on cluster membership}
We do not have definitive proof that our variable stars are members of 47 Tuc, the background SMC, or the Galactic halo. Spectroscopic observations are required to confirm membership; the heliocentric radial velocity of 47 Tuc (-19.4km/s; \citealt{Rich87}) is very different to that of the Galactic halo and the SMC (+158km/s; \citealt{May84}). In this paper, we have estimated membership for variables from their location on the cluster Colour Magnitude Diagram (CMD), which is shown schematically in Fig.\ref{linecmd}. As a globular cluster, 47 Tucanae has a very small spread of stellar metallicity, so that the stars lie on tight loci on the cluster CMD, without the spread in colour observed in other objects such as the SMC. As such, we can preliminarily assign memberships which are presented in Tables 1-5. For the contact eclipsing binary (EcB) systems, we have adopted Rucinski's absolute brightness calibration \citep{Ruc93} to calculate M$_V$, distances and thus memberships. The method estimates M$_V$ from the period, unreddened colour, and system metallicity via:

\begin{center}
$M_{V}^{cal}=-4.43\log(P) + 3.63(V-I)_O - 0.31 - 0.12 \times \rm{[Fe/H]}.$
\end{center}

For all our EcB systems, we have adopted [Fe/H]=-0.76 and E(V-I)=0.05 \citep{Harris96}. Fig.\ref{Ruccal} shows the derived distance modulus of all EcB for which we have complete colour information and have periods P$<$1d; such systems can be regarded as contact systems. The derived distance modulus (DM) of 47 Tuc from this plot is 13.14$\pm$0.25, which is in agreement with the estimate of 13.21 presented by \citet{Harris96}, and that derived by \citet{Per2002}, using main sequence fitting, of $(m-M)_V = 13.37^{0.10}_{-0.11}$. From Fig.\ref{Ruccal}, it is clear that we have detected 10 eclipsing binaries that are likely members of 47 Tuc. V95 and V26 appear to be foreground members of the Galactic halo, whereas V20 is a likely member of the Small Magellanic Cloud \citep{Harries03}. Interestingly, V11 and V75 both have distance moduli which lie inbetween 47 Tuc and the SMC.  It is interesting to note the very small amplitude of variation associated with the foreground EcB V26, perhaps indicating low mass components.

To estimate the total number of variable stars present in the WFI field we consider how many of \citep{Kal98}'s sample we missed due to telescope offsets and gaps between our CCDs. We recovered 31 of the 42 variables presented in that paper. Using this result, we estimate that we missed $\sim$26$\%$ of the variables we are capable of detecting. We therefore expect that in total there are 126$\pm$11 detectable variable stars present in the field.

\subsection{RR Lyraes}
A significant number of Small Magellanic Cloud (SMC) RR Lyrae stars were found in our data. Their phase-wrapped lightcurves are presented in Figs.\ref{VarPlot4}-\ref{VarPlot7}, along with a preliminary examination. All but four RR Lyrae stars are clustered around V=19.68, indicating their membership in the SMC. The other four stars are very likely to be located in the Galactic halo, one of which (OGLEGC223=V10) was presented in \citet{Kal98}.
Three populations of RR Lyrae stars are found in our data: type AB, type C, and two examples of RR Lyrae stars with periods of about 1 day. These latter stars (type AHB1) are described by \citet{Sand94}, and constitute very low metallicity post-horizontal branch (HB) stars passing rapidly through the instability strip in the vicinity of the Horizontal Branch but on bright evolutionary tracks \citep{Strom70}. Both of our AHB1 stars are Galactic Halo stars, and hence indeed likely to be of a low metallicity. The identification of these AHB1 stars is presented in the RR Lyrae period distribution (Fig.\ref{RRLyrP}), and are also identified on the RR Lyrae P-V diagram (Fig.\ref{RRLyrPV}). 

By comparing the density of RR Lyraes in our sample with those previously published for the field of 47 Tuc, we can draw conclusions about the completeness of our RR Lyrae sample.
\citet{Grah75} searched for variables in an area covering 4680 arcmins$^2$ north of 47 Tuc, which included a small part of the cluster. The RR Lyrae density presented in that paper was 0.016 variables per square arcminute, and is the generally accepted value for the RR Lyrae density in this region of the sky. 
The surface area covered by our search is 2704 arcmin$^2$, and yields a derived RR Lyrae density value $\thickapprox$ 0.016$\pm$0.003 RRLyr/arcmin$^2$, consistent with the \citet{Grah75} result. The limiting magnitude of Graham's search was B$\thickapprox$20.3, whereas ours is V+R$\thickapprox$21.0. We did not identify any 47 Tuc RR Lyraes in our dataset, due to their relative brightness compared to our target stars. With our 300s exposures, at V=14.06 \citep{Leon2000} such cluster stars would be very close to saturation.

Fig.\ref{RRLyrPV} shows the period-luminosity diagram for our sample of RR Lyraes. Those that lie in the SMC are easily distinguishable from those that lie in the Galactic Halo (marked with a H). The long-period AHB1 RR Lyraes are also identified. The average V magnitude of the SMC stars lies at V=19.68$\pm$0.24. The mean V absolute magnitude of RR Lyraes in the SMC is M$_{V}$(RR)=0.75, as assumed by \citet{Grah75}, and also used by \citet{Kal98} and includes correction for metallicity effects. It therefore follows that the SMC distance modulus (m-M)$_V$ from our sample of RR Lyraes is 18.93$\pm$0.24 or 61.45$\pm$7.0 Kpc. The large errorbar is due to the relatively small sample size. The average magnitude of each of the RR Lyrae stars was found, by integrating across the lightcurve, and then the phase of the variability amplitude at the time of the CMD dataset was used to determine the actual apparent magnitude at this same time. The difference between this and the mean magnitude was then measured. The error in this measurement as plotted on Fig.\ref{RRLyrPV} was taken as the residual scatter in the phase-wrapped lightcurve points at this time. This method allows us to account for the variability amplitude when finding the V magnitudes of our stars, and hence a more accurate measure of the SMC distance. Our result compares favourably with that of 18.89$\pm$0.10 presented by \citet{Harries03}. A larger sample size would help to indicate more conclusively if the SMC is extended in the radial dimension towards the direction of 47 Tuc.

The Blazhko Effect is a little understood feature of some RR Lyrae stars, in which the amplitude of variation itself varies with a certain periodicity. One popular theory to explain this erratic behaviour is that it is related to the presence of a strong photospheric magnetic field \citep{Cous83}, yet recently \citet{Chad2004} has ruled out a magnetic field for the brightest Blazhko (BL) star - RR Lyrae itself, the prototype of the class. Neither this, nor the rotating resonant pulsator model \citep{Dziem99} explain observed BL star properties. The frequency of such BL stars seems to be dependent on the metallicity of the environment in which they occur, \citet{Alcock03} has found a different incident rate of BL stars in the LMC ($\sim$11$\%$) when compared to \citet{Mosk03}, who found a 23$\%$ incidence rate for the Galactic Bulge. Both papers agree that metallicity is the most probable reason. The reason for BL behaviour in general therefore remains unknown. It is seen on a few of our RR Lyrae sample (V2, V3, V18, V86). Fig.\ref{V86B} shows the remarkable regularity of the Blahzko Effect for V86, for which we plot the phase-wrapped lightcurve over four periods. The amplitude of the variation decreases significantly over three primary periods. This behaviour was constant over the whole 33-night run. The other three Blahzko examples do not show such regularity. A few of the RR Lyrae lightcurves show significant scatter, primarily due to crowding, and variable seeing. 

\subsection{Long Period Variables}
Our sample of variables also includes 20 Long Period Variables (LPVs) with locations on the CMD that are consistent with two main populations: those present in the red giant branch (RGB) of 47 Tuc itself, and those which lie on the AGB/RGB of the SMC. We define a long period as being significantly longer than that of RR Lyrae and EcB stars, from about 5-6 days. Stars with very long periods cannot have their periods accurately determined from our dataset, as our temporal coverage is insufficient to show even a single period; we present lower limits on the periods of the more extreme LPVs. Some of these stars were found by \citet{Kal98}. These extreme LPVs are likely to be examples of AGB stars (Miras) in the SMC. V27 (OGLEGC230) is affected by bad seeing and extreme crowding, and hence has significantly more scatter on its lightcurve. Our derived period for this star is half that derived by \citet{Kal98}, who indicated that their period might require revision.

Only about six of our LPV sample can be given tentative 47 Tuc membership, a number too small to give meaningful results to study their radial distributions. It is very likely that the majority of cluster RGB variables are brighter than our magnitude limit. A shorter exposure search for variability among 47 Tuc RGB stars has recently been started by Kiss et al (2003, private communication, also with WFI on the MSSSO 40-inch), and our LPV sample will overlap somewhat with their results. This should allow a more accurate study into the cluster LPV radial distribution.  

\subsection{Other Variables}
A small number of other variables were also discovered in our dataset, including two Cepheids, four $\delta$ Scuti stars, and an anomalous short-period red variable, which is a likely SMC star. The two Cepheids (V24 and V37) are identified from their position in the schematic cluster CMD (Fig.\ref{linecmd}). They are significantly brighter and redder than the RR Lyrae stars, but are of short period (0.387d and 2.572d respectively) for Cepheids, and as such could be classified as anomalous. V24 has been tentatively identified as a TypeII Cepheid based on the secondary variation seen on the lightcurve at phase $\sim$0.5.

We detected four $\delta$ Scuti stars in our search (V35, V54, V67 and V80). All are certainly members of the SMC, as they have V$\sim$21. This is at the limit of our detectability, but they were found due to their large amplitude of variation. V35,V54 and V67 all have very short periods $<$0.1d, typical of $\delta$ Scuti stars. V80 has a period that is longer, at 0.2144d, and is among the faintest variables in our catalogue at V$\sim$22. It has been classified as a $\delta$ Scuti star due to the shape of the lightcurve, and the amplitude of variation. The $\delta$ Scuti lightcurves presented here show a significant amount of scatter, which is attributed to photometric scatter caused by the faintness of the sample. To investigate the possibility of multi-periodicity, the periodograms for these four stars were compared to those of non-variables of the same magnitude. The lightcurves were phase-wrapped to all significant periods found with the LSP (see earlier), the majority of the extra periodicities were found to be common, and attributed to systematic effects. No obvious secondary period was seen for these stars down to our sensitivity level. The amount of photometric scatter at V$\sim$21 is $\sim$0.3 mag, and is consistent with the scatter seen on these four lightcurves.

Finally, from its position on the schematic CMD V60 is an apparent SMC AGB star, with V-I=2.99, yet a periodicity of only 0.2544d. The amplitude of variation is moderate ($\Delta$mag$\sim$0.1), and as such indicates an apparent giant red star with a very short period pulsation. It follows that such a star would likely be unstable at such short periods. At the distance of the SMC derived above, the absolute magnitude would be -1.8, typical of an A-type star. It therefore seems likely that this star is actually much bluer, but could possibly be a post-AGB star surrounded by a dusty shell. Future monitoring and imaging of this star would be useful to unravel the mystery.   

\section{Summary and Conclusions}
We have presented data for 100 variable stars detected across a wide (52$\times$52$'$) field centered on the globular cluster 47 Tucanae. Of these 100, 69 are new discoveries. The sample consists of 41 apparent Small Magellanic Cloud RR Lyrae stars, four Halo RR Lyrae stars, 28 eclipsing binaries, 20 Long Period Variables, four $\delta$ Scuti Stars and two Cepheids. We also detected one anomalous short period red giant, perhaps surrounded by a dusty region. Four of our RR Lyrae sample display Blahzko Effect variations, one with remarkable regularity. This catalogue more than doubles the number of known variables in the 47 Tuc field. Of the EcB sample, four variables are perhaps orbited by faint companions, most likely M-Dwarf stars. Such stars are important in determining the long-term stability and survivability of low-mass objects in close orbits inside globular clusters. Future spectroscopic observations of these candidates are planned. As well as presenting this new variable catalogue, this paper presents a new complete database of V and I photometry, astrometry and 33-night V+R lightcurves for 109,866 stars across the field. The distance modulus of both 47 Tucanae and the SMC have been determined from our sample. The values are consistent with those already published for these two objects.

It is clear from the eclipsing binary results, as well as those previously published in the literature, that the relative frequency of contact binaries in the field of 47 Tuc is very low compared to other studied regions. The reasons for this difference are unclear, although mass-segregation and dynamical effects seem to play an important role. Our sample of EcB shows a distinct period/radial-distance segregation, perhaps indicative of dynamical relaxation. Quite possibly the more massive shorter period contact binaries are located preferentially within the cluster core, an unsampled area in our experiment. 

\section*{Acknowledgments}
The authors wish to thank the following people for their help in the producing this piece of work:
 Matthew Coleman for helping the first author relieve the insanity of observing for 33 nights straigh; Eduard Westra for help in debugging the photometric pipeline, and for answering many emails to that effect; Brian Schmidt for allowing use of his astrometric package, as well as instructions on its operation; Laszlo Kiss for looking over some of the lightcurves and clearing up the identifications of the more noisy variables, and Doug Welch for helpful comments while acting as referee.

Finding charts for all variables are available on the Electronic edition of this paper. The orientation of these charts has north to the top and east to the left, produced from the corresponding template image which contains the variable, and they are each 3.24$'$ on a side.

\clearpage

\plotone{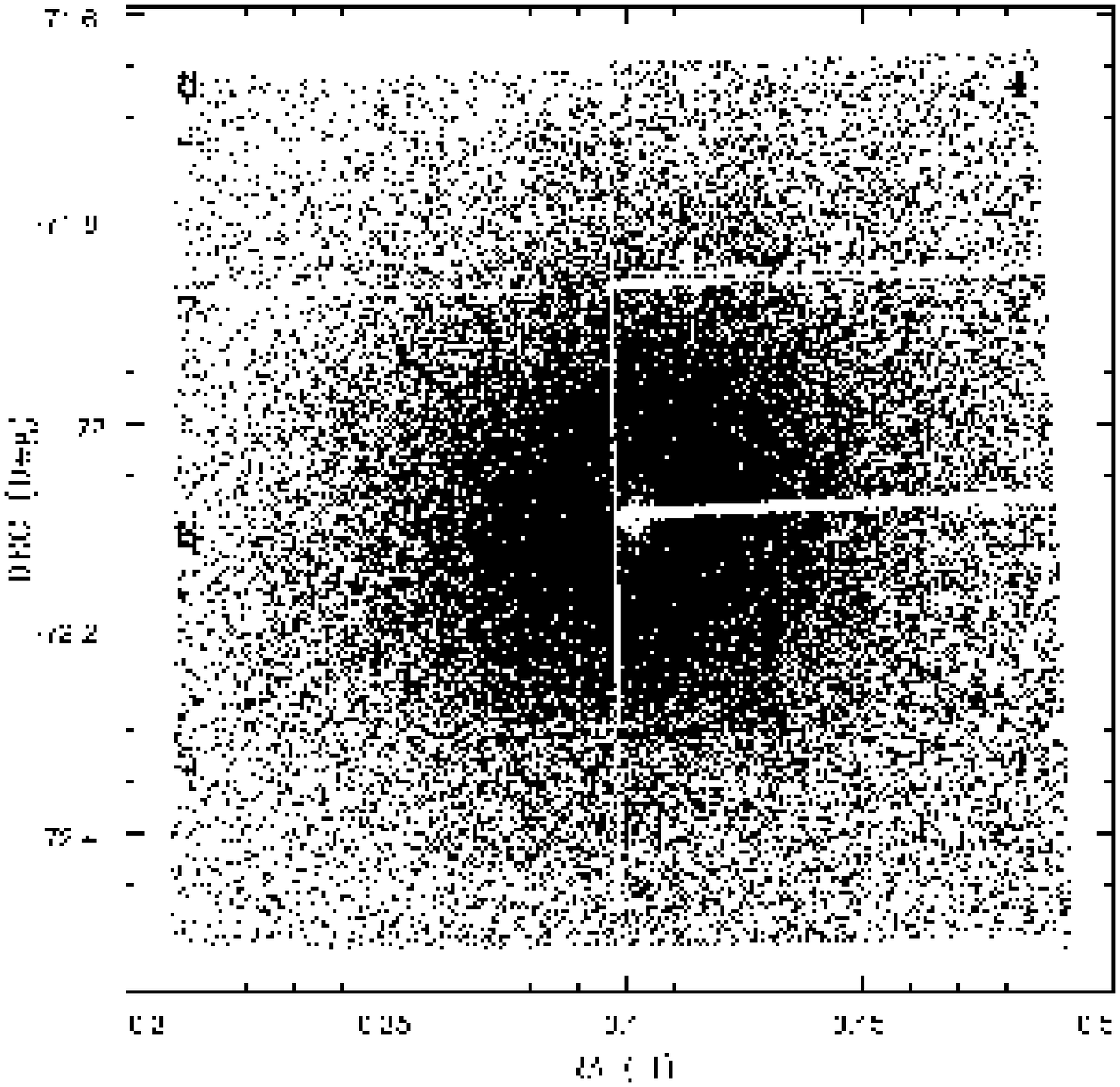}
\figcaption[Weldrake.fig1.eps]{Astrometry and layout of the 47 Tuc WFI field; detected variable stars are plotted as triangles. The eight CCDs are labelled numerically. The two regions used to produce the RMS error plot of Fig.\ref{rmstot} are indicated as boxes in CCD3 and CCD4. The HST field (Gilliland et al 2000) is marked with a bold box, indicating the significant increase in our field of view compared to those observations.The arrow shows the direction to the core of the SMC.\label{astromplot}}

\plotone{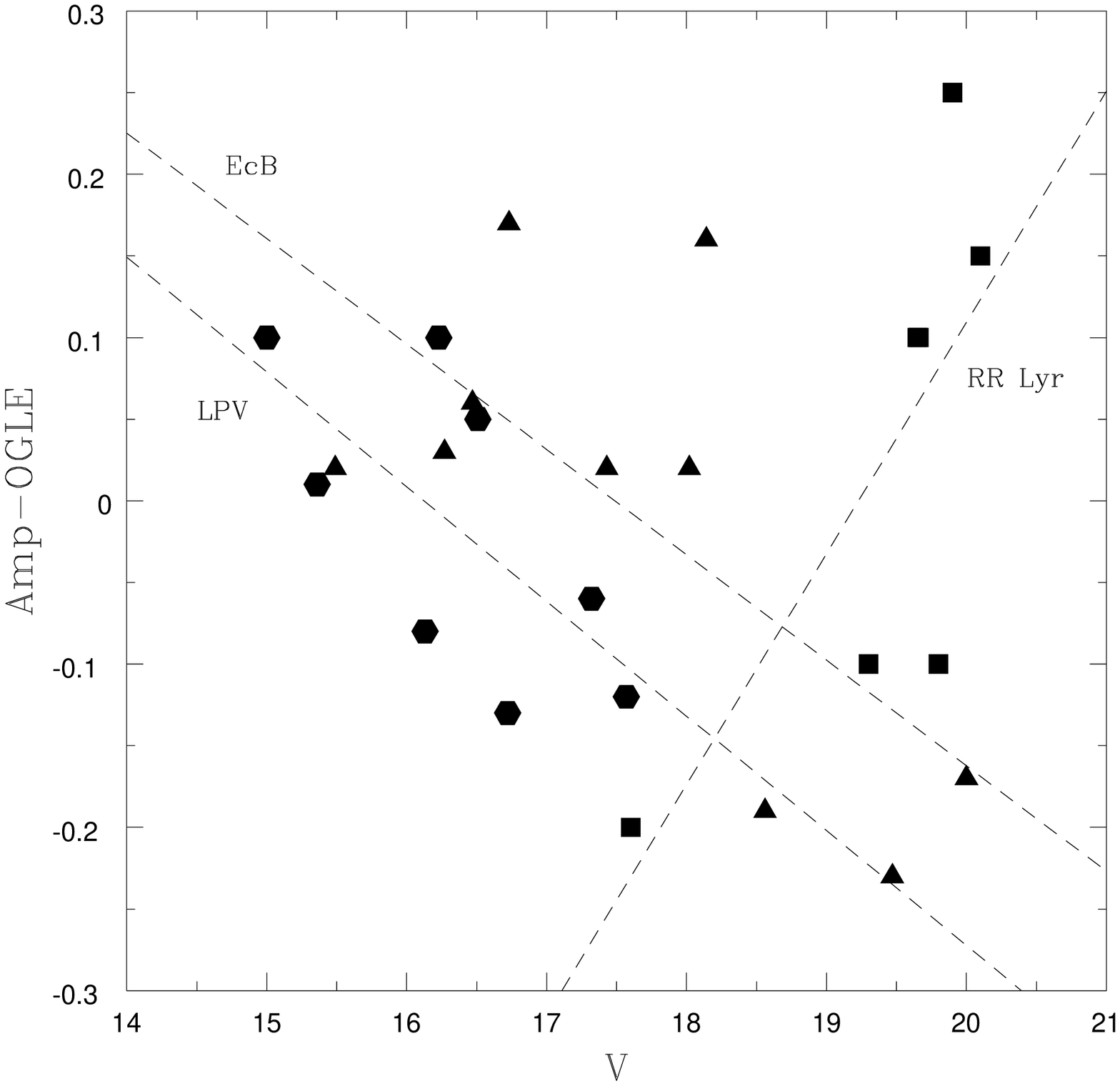}
\figcaption[Weldrake.fig2.eps]{A comparison between the total ampltudes of our variables and those cross-identified with \citet{Kal98}. The LPV (hexagons), EcB (triangles) and RR Lyrae (boxes) variables have been plotted and the least-squares fit through the data has been added. If our amplitude is larger than the comparison, the value plotted is $>$0. The trends are attributed to the different passbands used (LPV), the presence of different coloured components (EcB) and to the faintness of the sample (RR Lyraes).\label{ampmags}}

\plotone{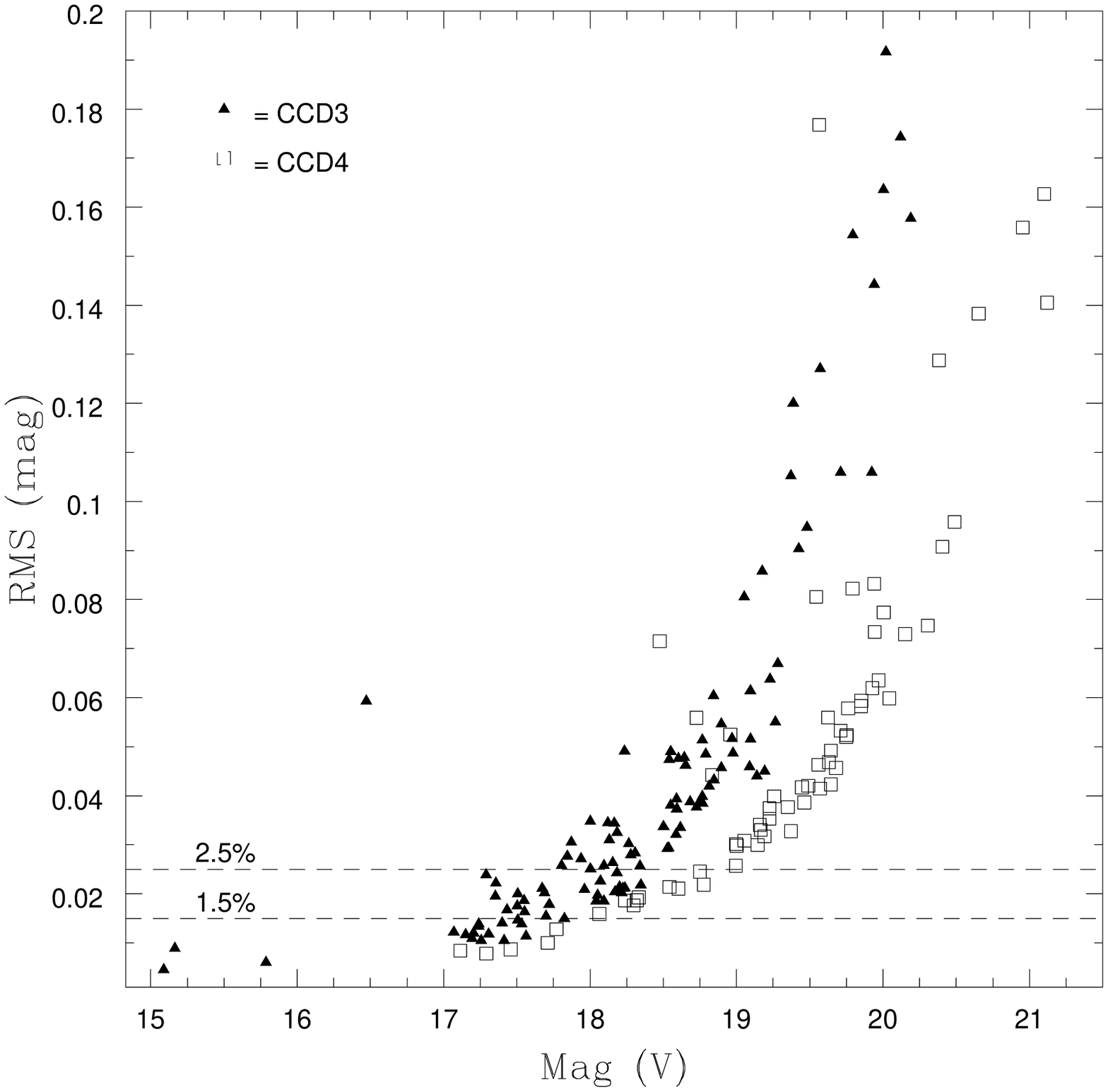}
\figcaption[Weldrake.fig3.eps]{Photometric precision, measured as RMS uncertainty  for both an inner crowded CCD (triangles) and an outer uncrowded, CCD (squares). At V$\leqslant$18.5, the quality of the photometry is sufficient in the four inner CCDs to detect a $\sim$1.5-2.5$\%$ dip typical of that caused by an orbiting 'Hot Jupiter' planet or brown dwarf, and is of comparable quality to V$\leqslant$19.0 in the outer four uncrowded CCDs.\label{rmstot}}

\plotone{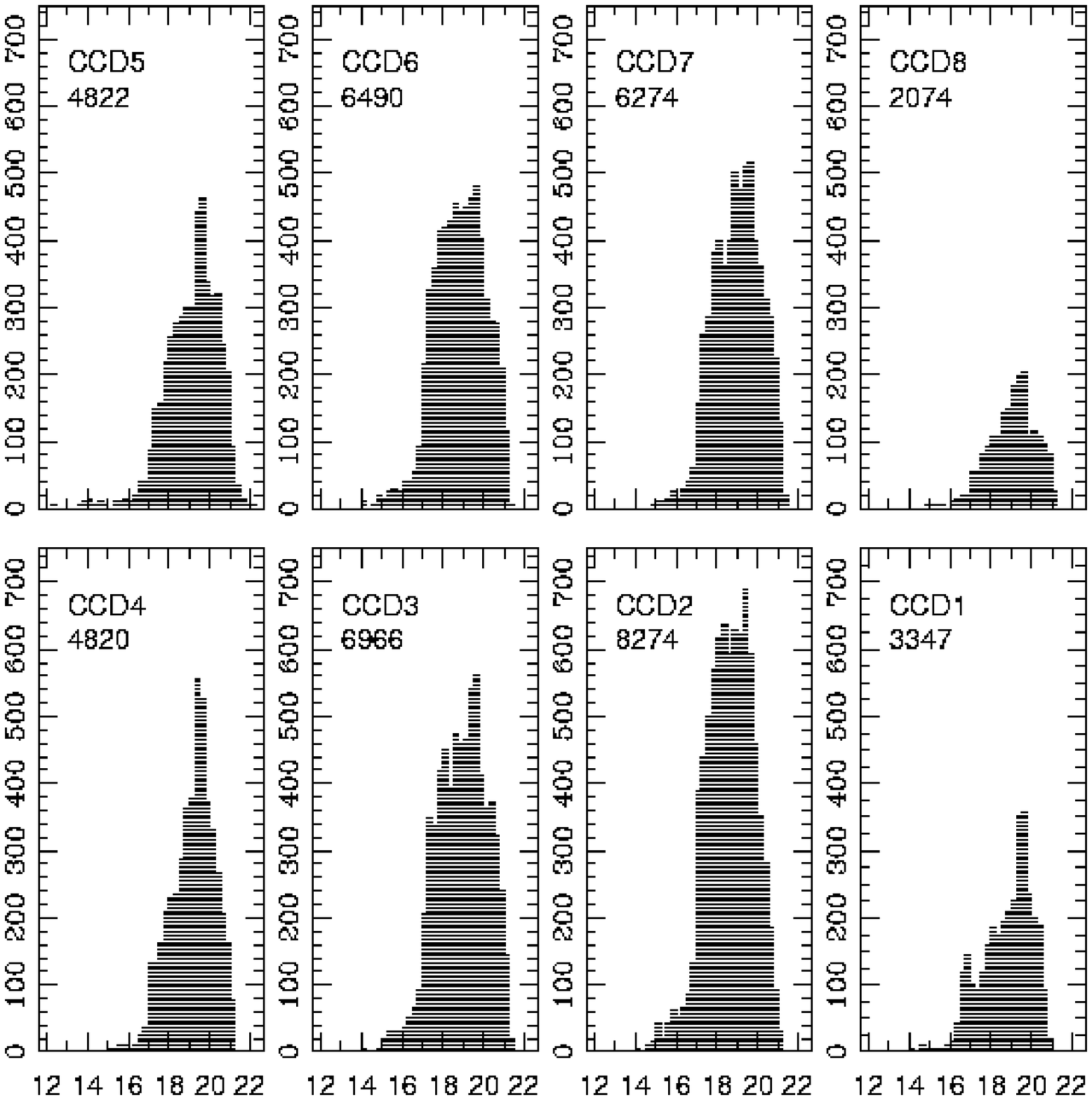}
\figcaption[Weldrake.fig4.eps]{As an indication of the depth of our V and I photometric dataset, the number of stars is plotted against magnitude for all eight WFI CCD's,as identified in Fig.\ref{astromplot}. The dataset is truncated at V=21, and is saturated at V=15 for CCD4 and V=13.5 for CCD5. The number of stars detected in each CCD is indicated at the top of each subframe.\label{starcount}}

\plotone{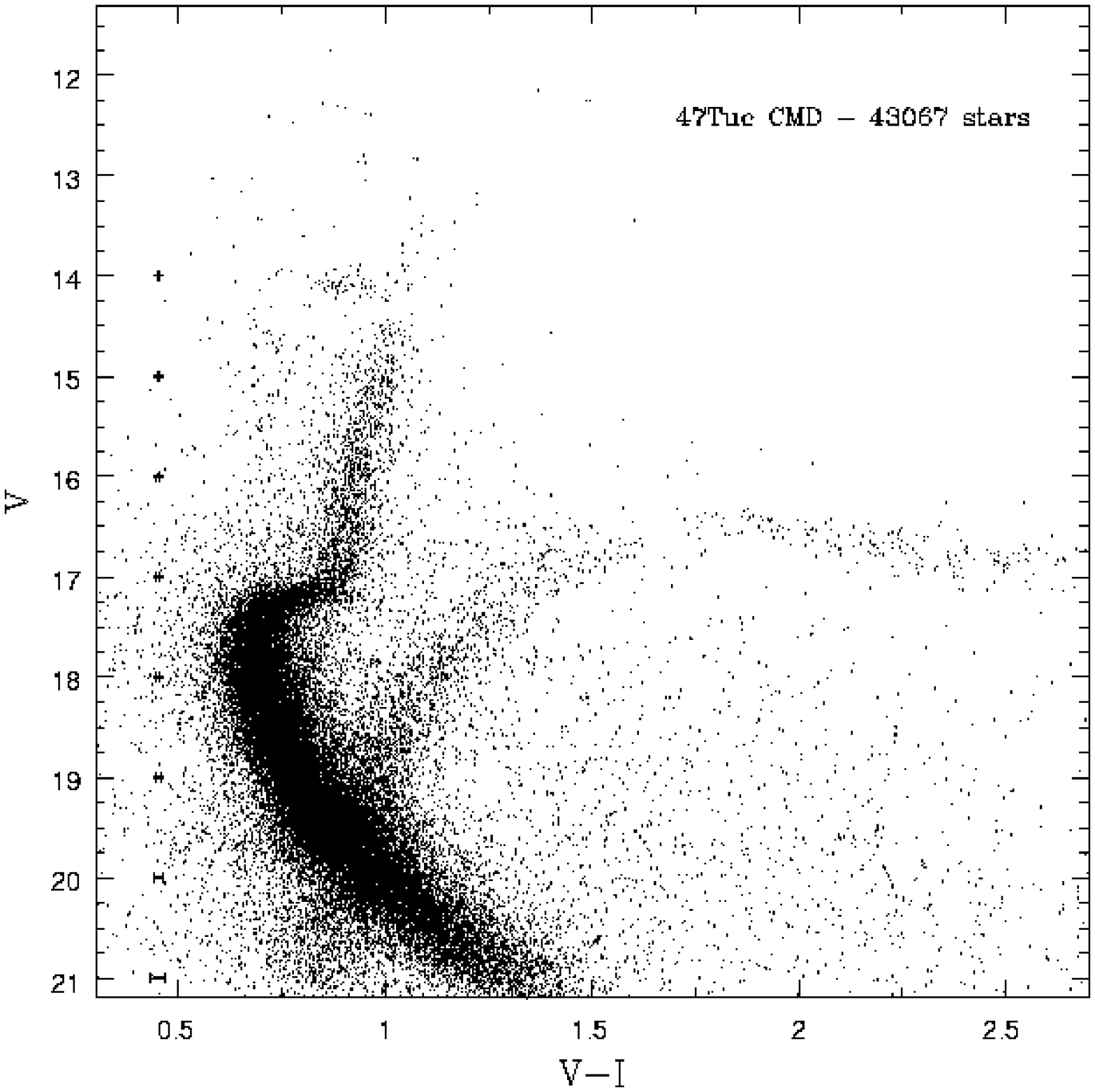}
\figcaption[Weldrake.fig5.eps]{Colour Magnitude Diagram dataset used to produce the colour information of the variable stars. The DAOPHOT output RMS errors in our photometry are plotted as errorbars. The calibration is accurate to better than 0.03 mag.\label{47TucCMD_total}}

\plotone{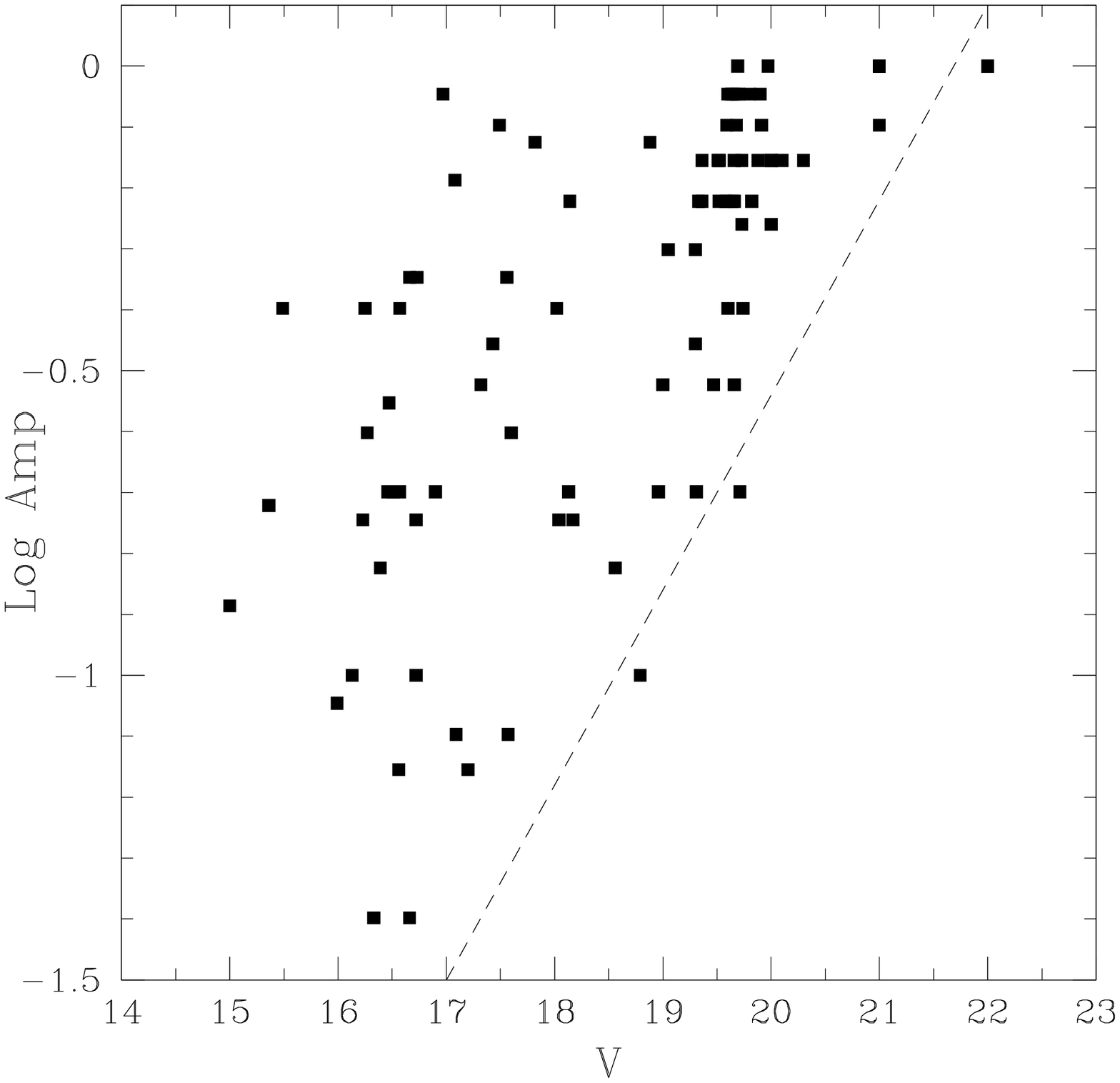}
\figcaption[Weldrake.fig6.eps]{The detection limit of our dataset. The dotted line indicates the minimum amplitude a variable must have to be detected as a function of V magnitude in our data. Any amplitudes $>$3$\%$ are detectable to a V of 17.0.\label{detlim}}

\plotone{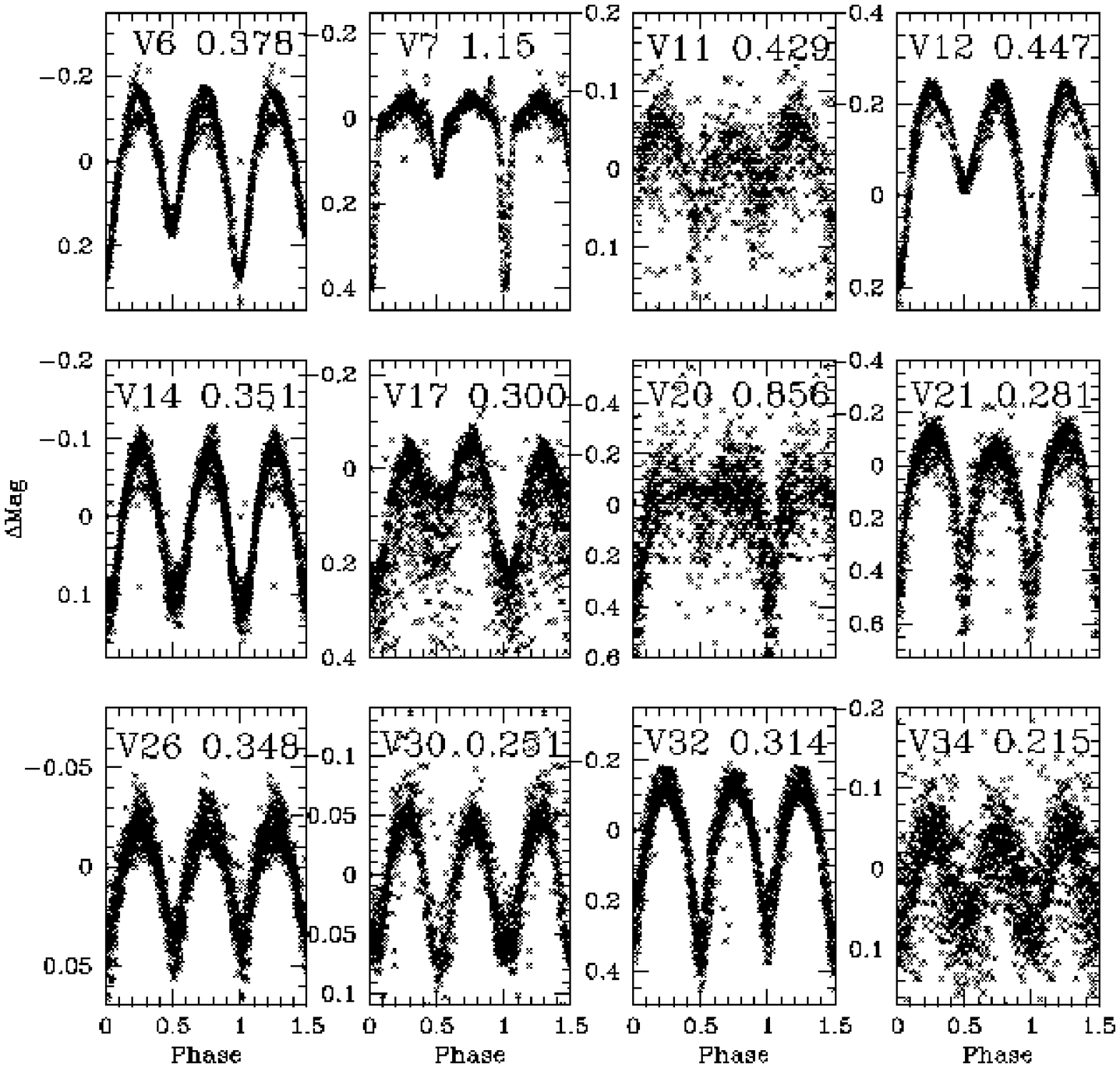}
\figcaption[Weldrake.fig7.eps]{Phase wrapped lightcurves of the Eclipsing Binaries detected in our dataset. The identification and period of each star is plotted together with the flux variation in magnitude units.\label{VarPlot1}}

\plotone{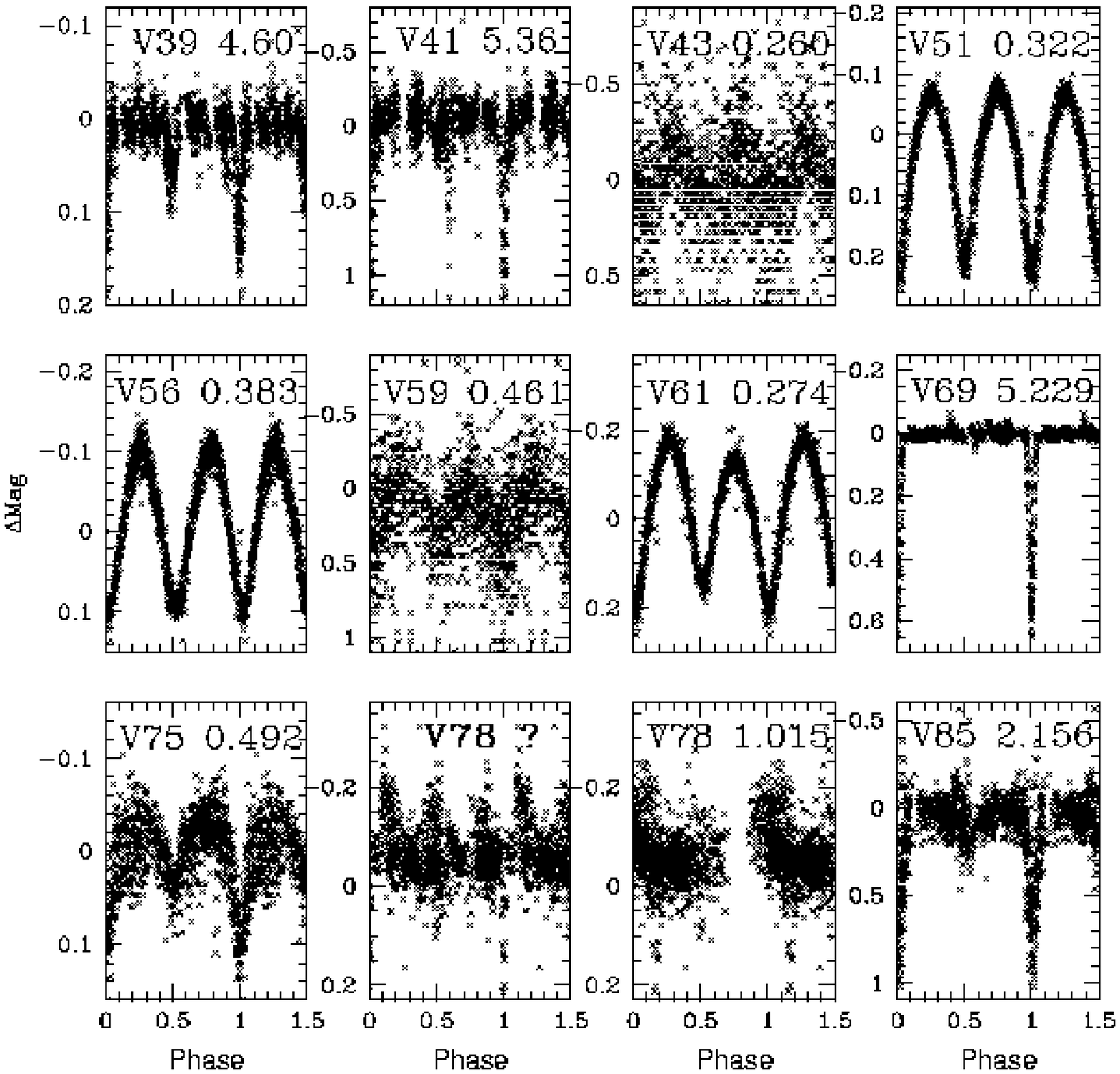}
\figcaption[Weldrake.fig8.eps]{Phase-wrapped EcB lightcurves (continued). V78 has an unknown period, as only one eclipse is seen over our sampling range. See Fig.\ref{Mdwarf}.\label{VarPlot2}}

\plotone{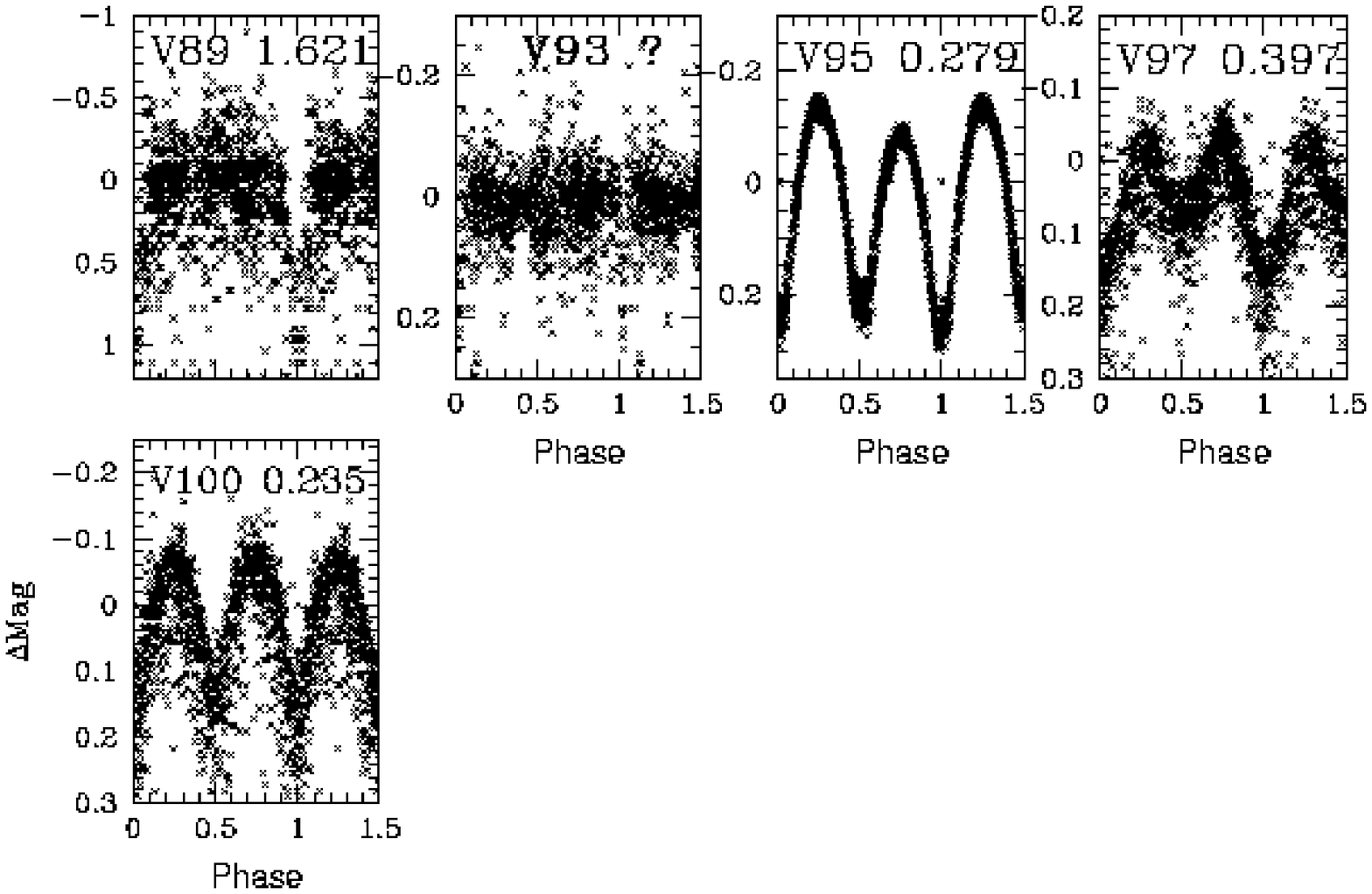}
\figcaption[Weldrake.fig9.eps]{Phase-wrapped EcB lightcurves (continued). V93 has an unknown period, as only one eclipse is seen over our sampling range. See Fig.\ref{Mdwarf}.\label{VarPlot3}}

\plotone{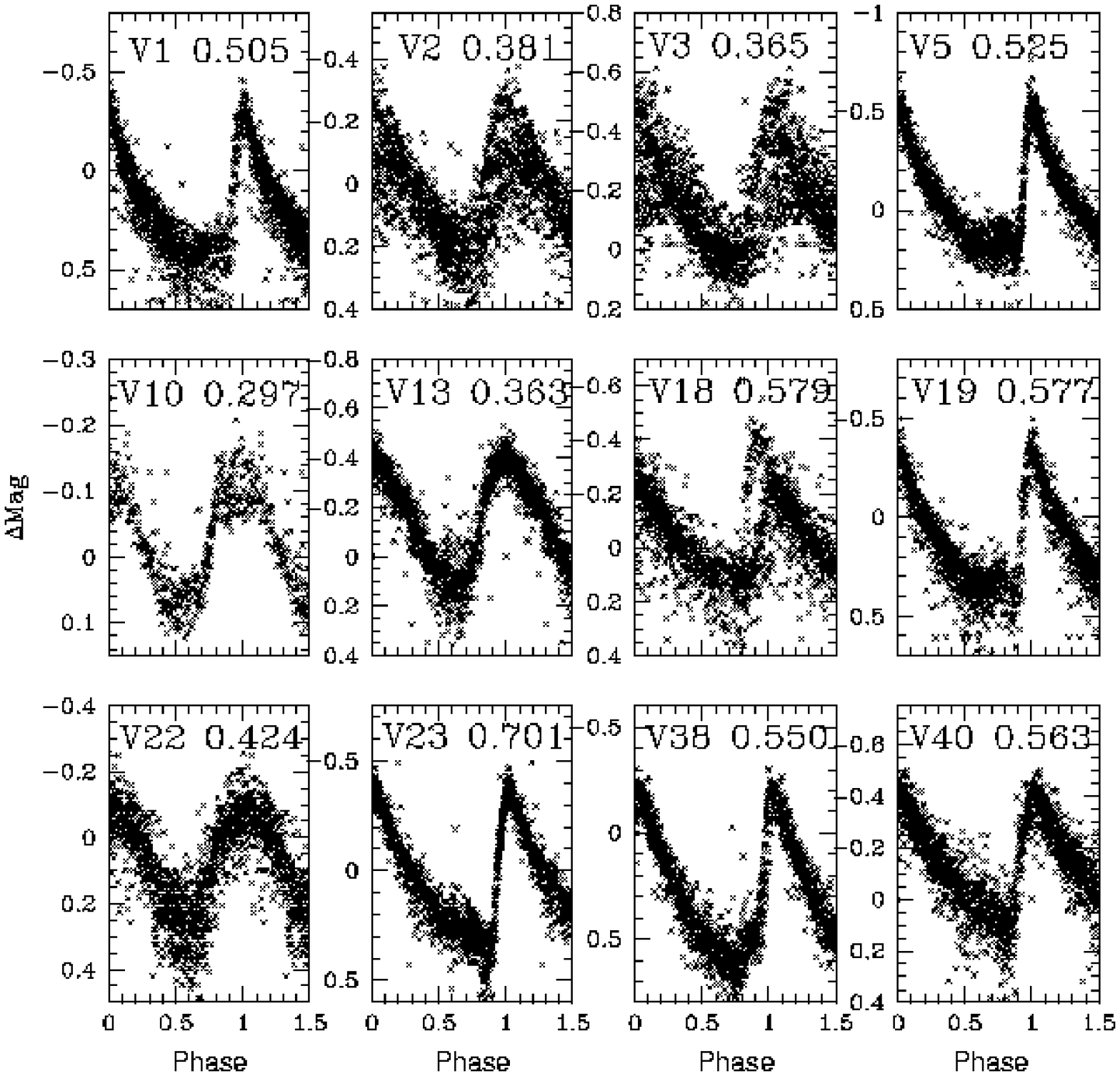}
\figcaption[Weldrake.fig10.eps]{Phase-wrapped lightcurves of the RR Lyraes detected in the dataset. The variable identification and the period is indicated.\label{VarPlot4}}

\plotone{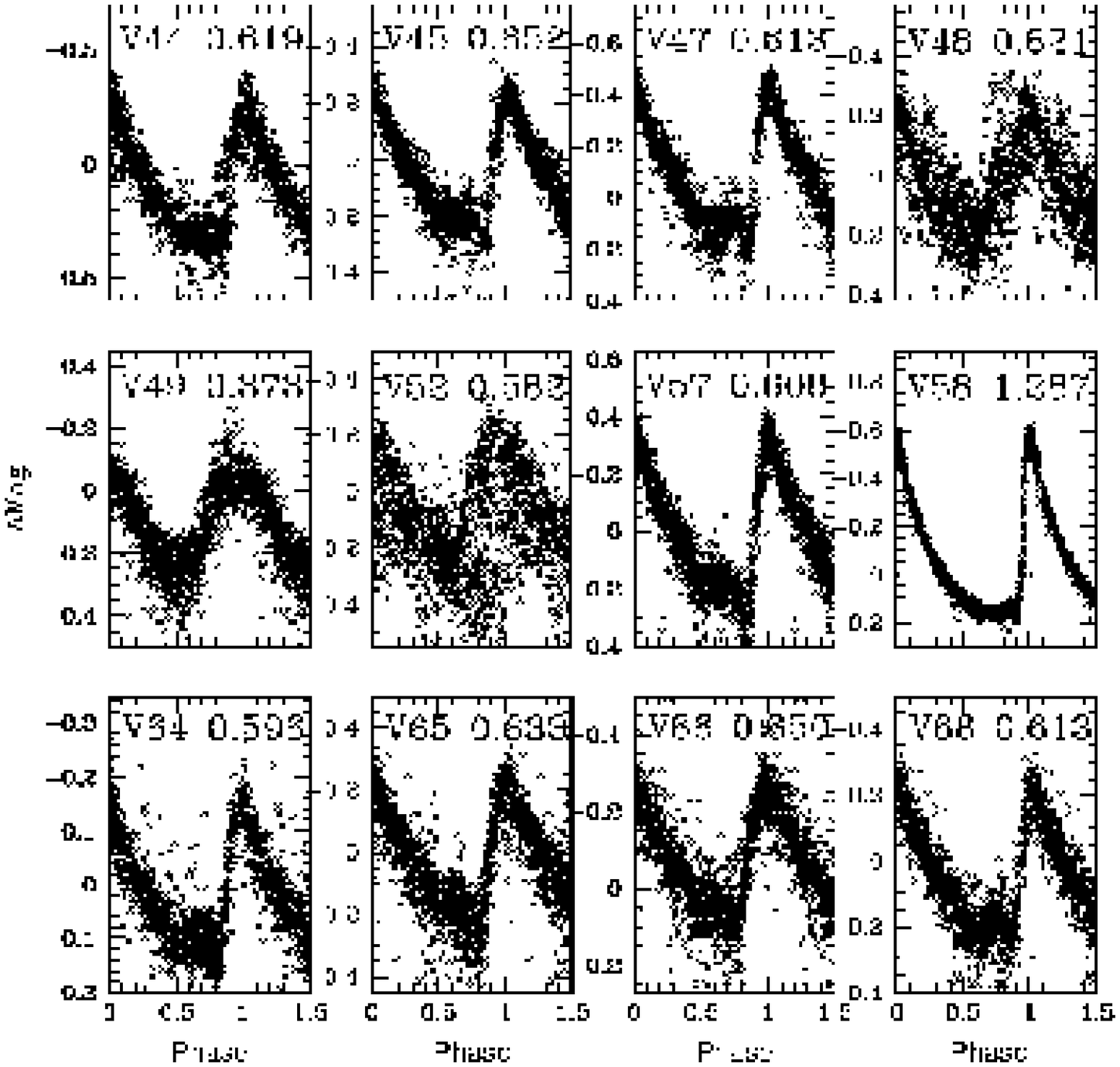}
\figcaption[Weldrake.fig11.eps]{Phase-wrapped RR Lyrae lightcurves (continued)\label{VarPlot5}}

\plotone{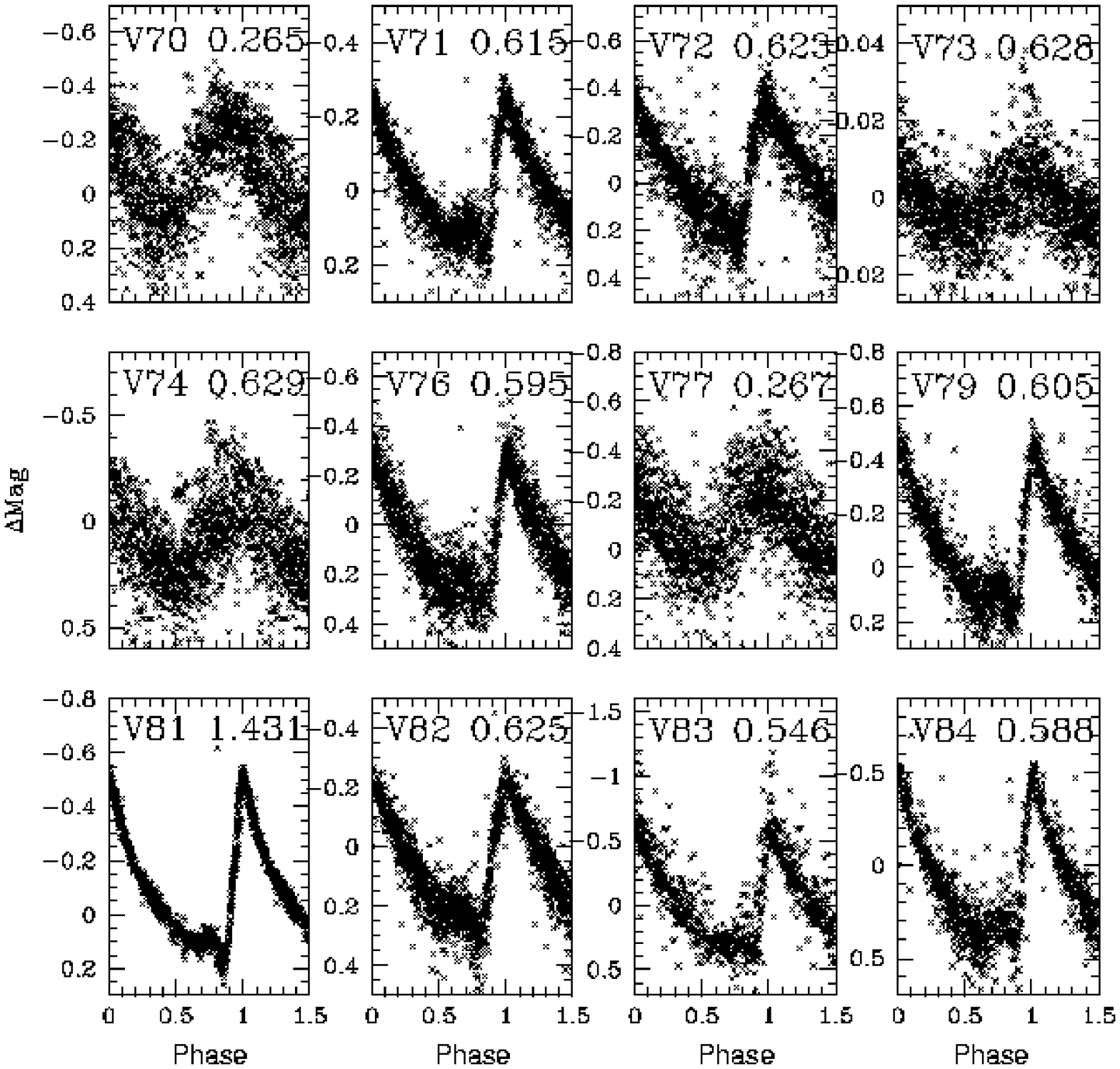}
\figcaption[Weldrake.fig12.eps]{Phase-wrapped RR Lyrae lightcurves (continued)\label{VarPlot6}}

\plotone{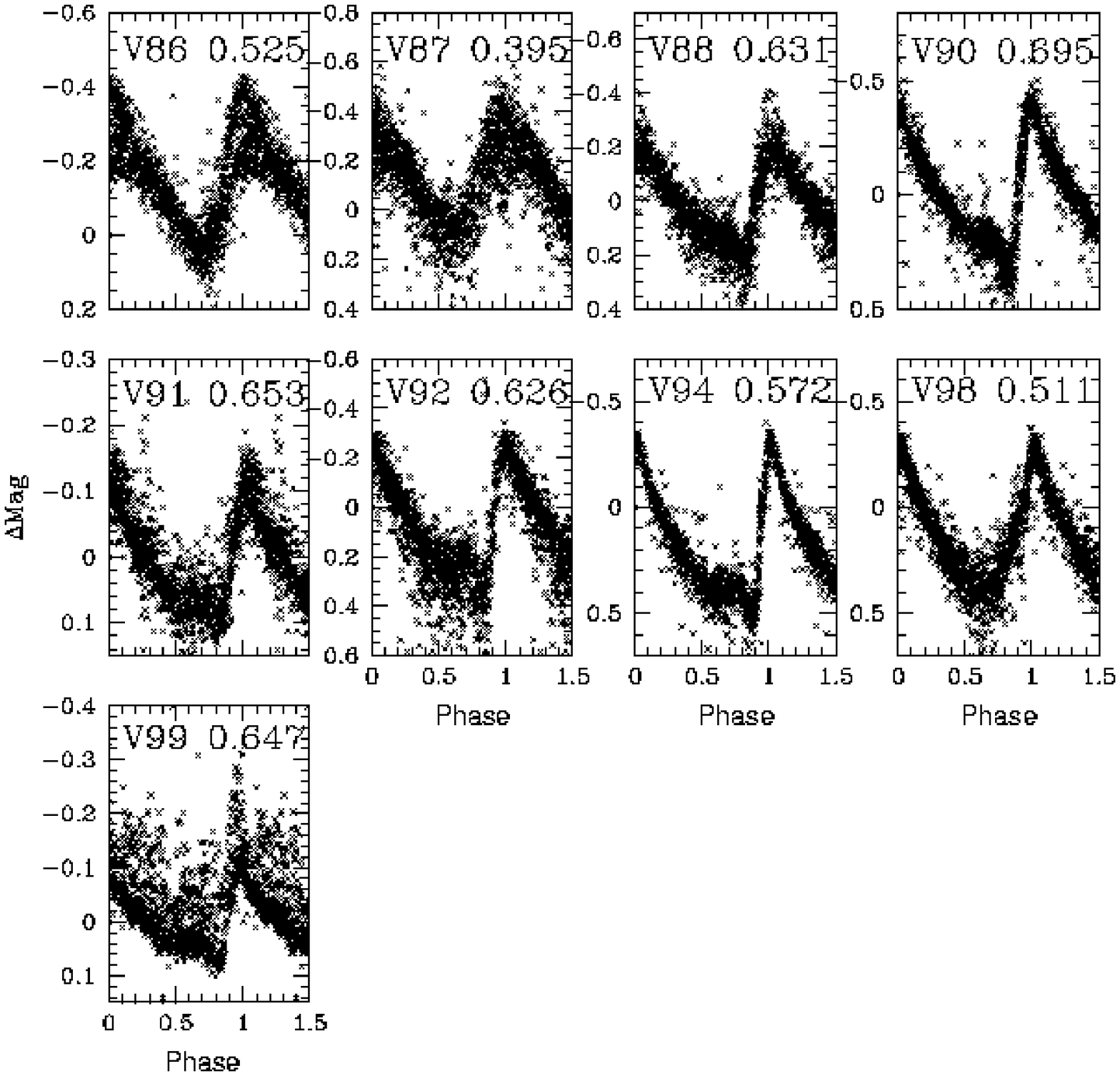}
\figcaption[Weldrake.fig13.eps]{Phase-wrapped RR Lyrae lightcurves (continued)\label{VarPlot7}}

\plotone{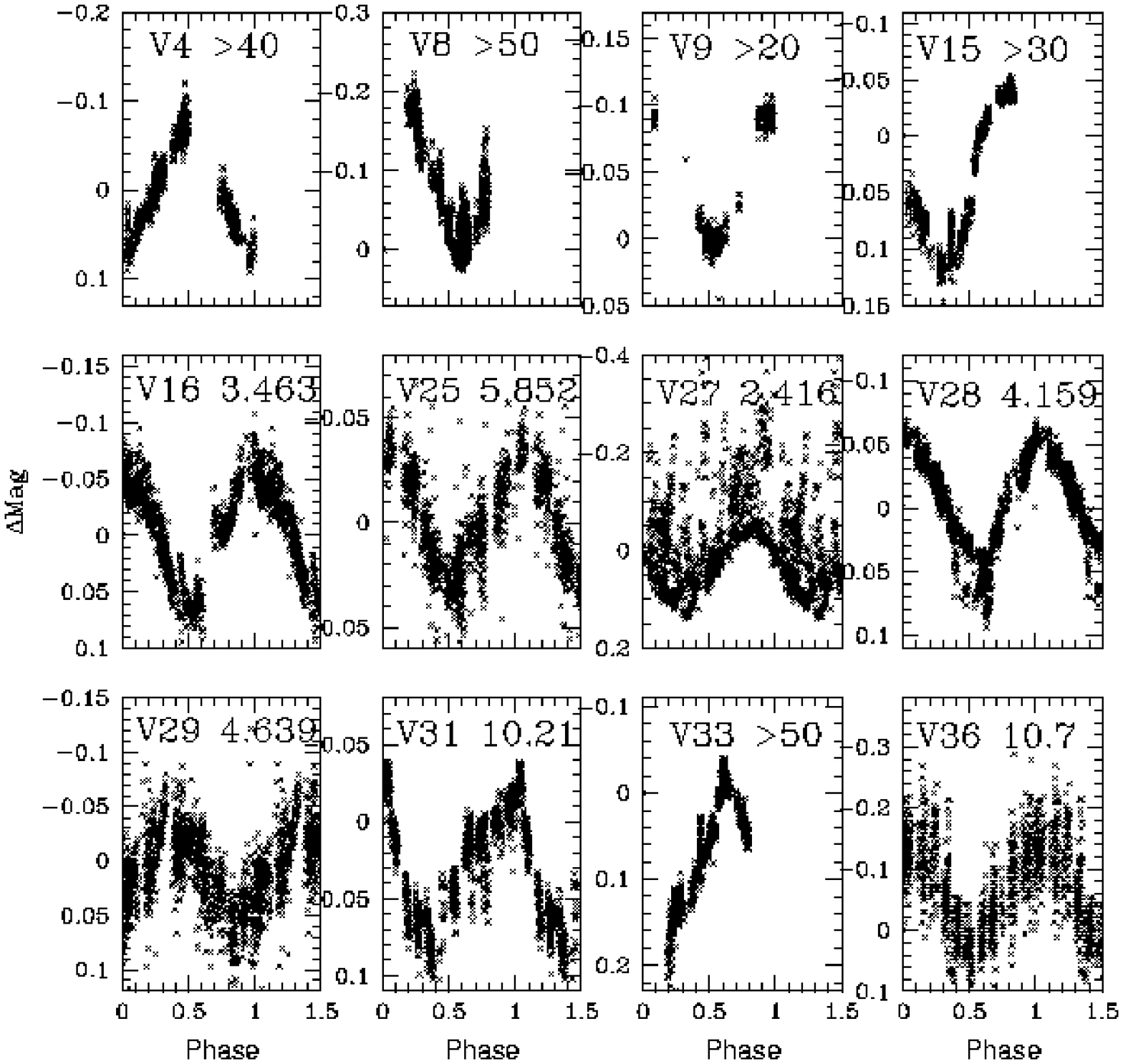}
\figcaption[Weldrake.fig14.eps]{Phase-wrapped lightcurves of the Long Period Variables (LPVs) detected in our dataset. The identification and period is indicated.\label{VarPlot8}}

\plotone{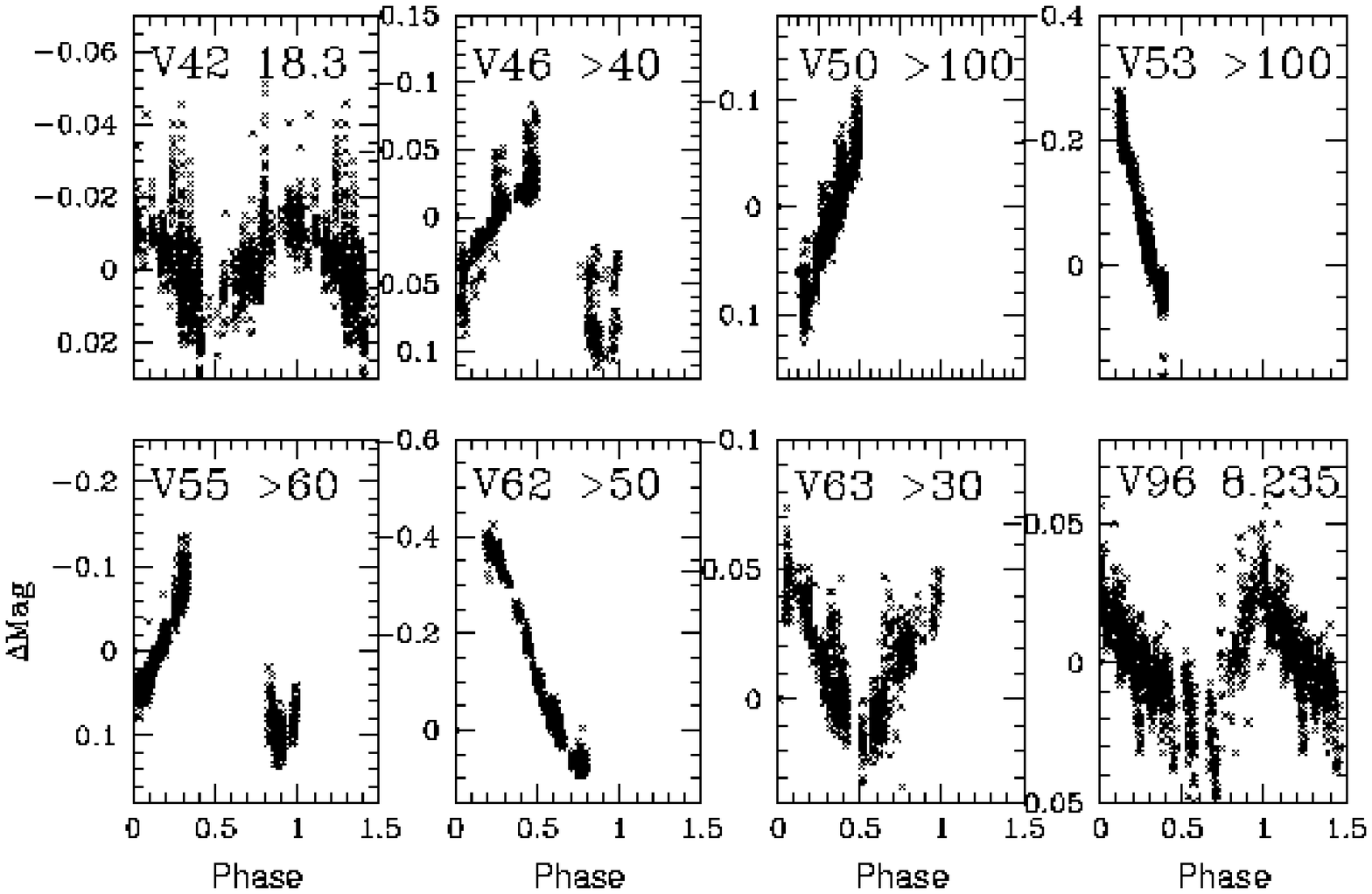}
\figcaption[Weldrake.fig15.eps]{Phase-wrapped LPV lightcurve (continued)\label{VarPlot9}}

\plotone{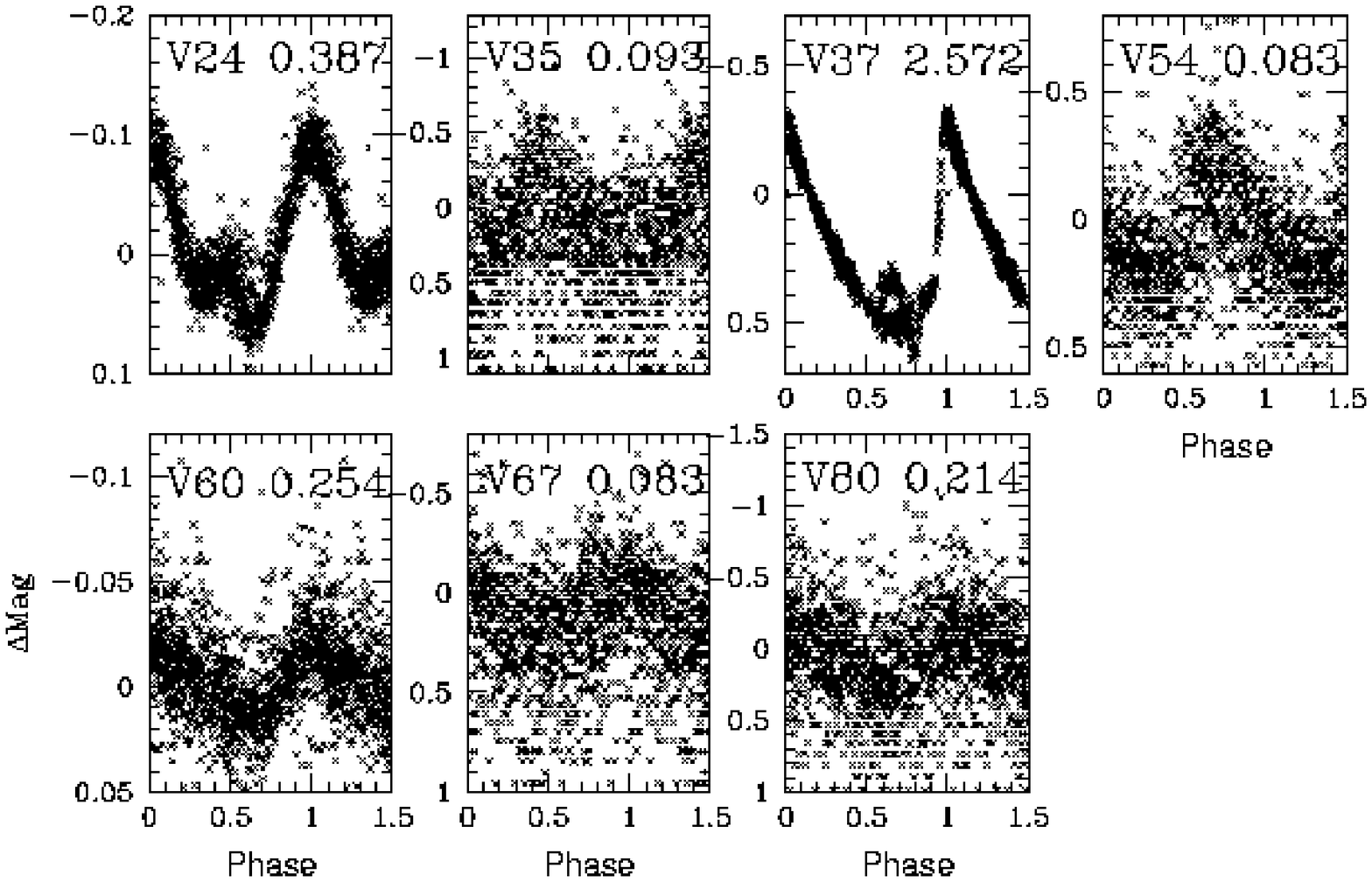}
\figcaption[Weldrake.fig16.eps]{Phase-wrapped lightcurves of the miscellaneous variables. The identification and period is indicated.\label{VarPlot10}}

\plotone{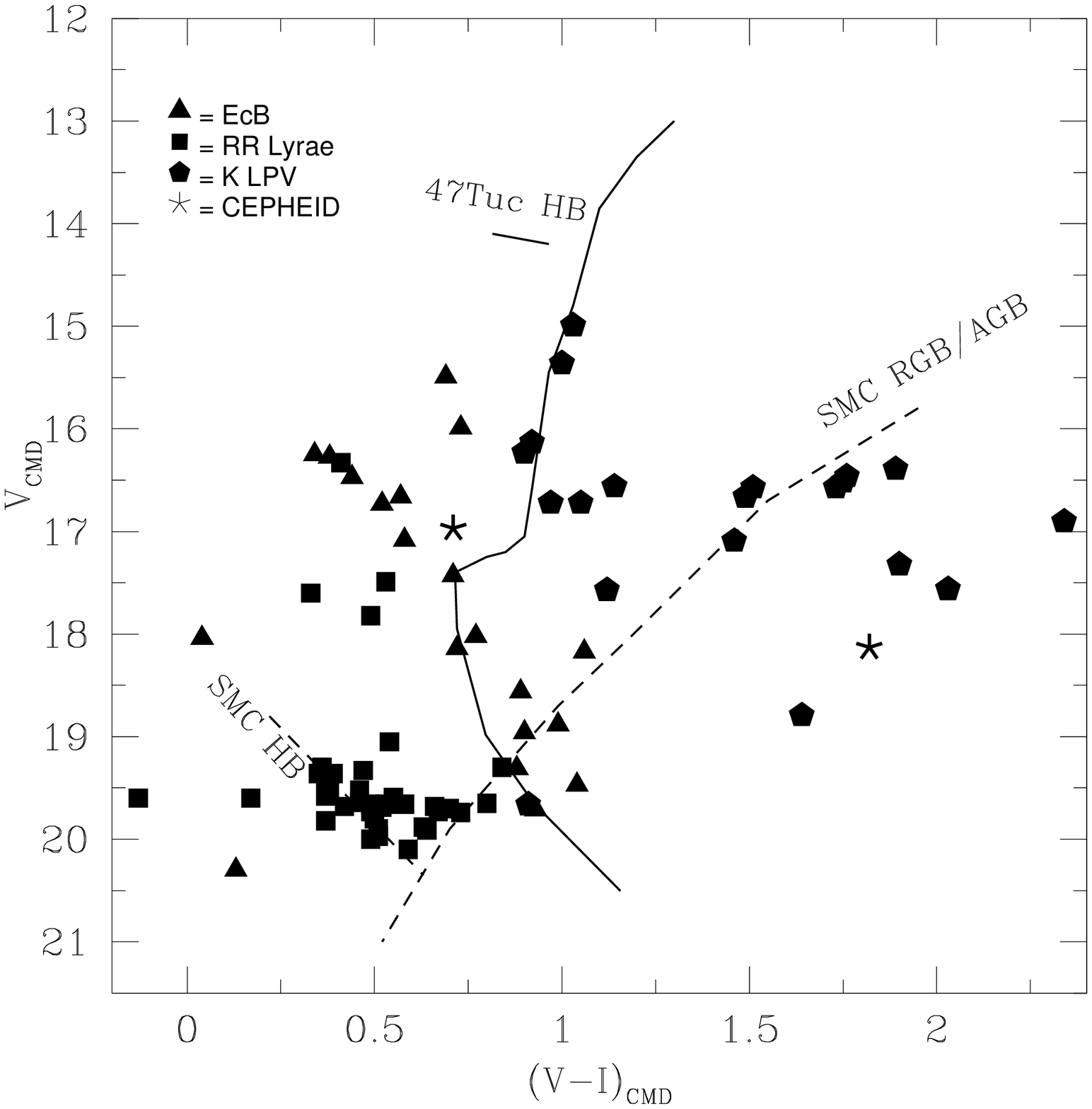}
\figcaption[Weldrake.fig17.eps]{Schematic Colour Magnitude Diagram of 47 Tuc, with the location of detected variable stars overplotted. The SMC Red Giant Branch and Horizontal Branch are also marked.\label{linecmd}}

\plotone{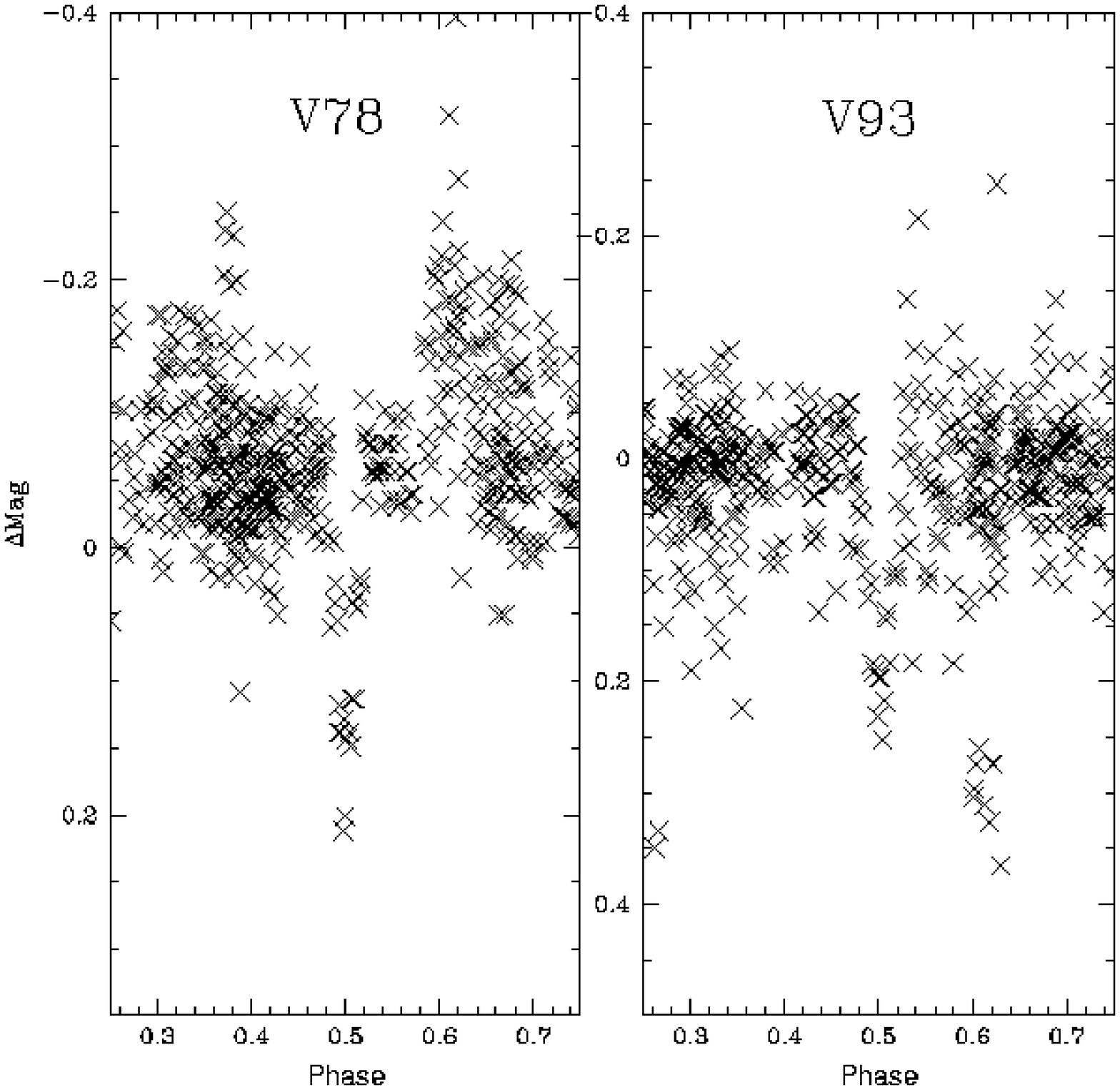}
\figcaption[Weldrake.fig18.eps]{V78 and V93, the two variables for which only one eclipse is seen, phase wrapped to the arbitrary values (see text) of 2.794d and 2.419d respectively, plotted to show more detail. The eclipse is likely caused by an M-Dwarf companion.\label{Mdwarf}}

\plotone{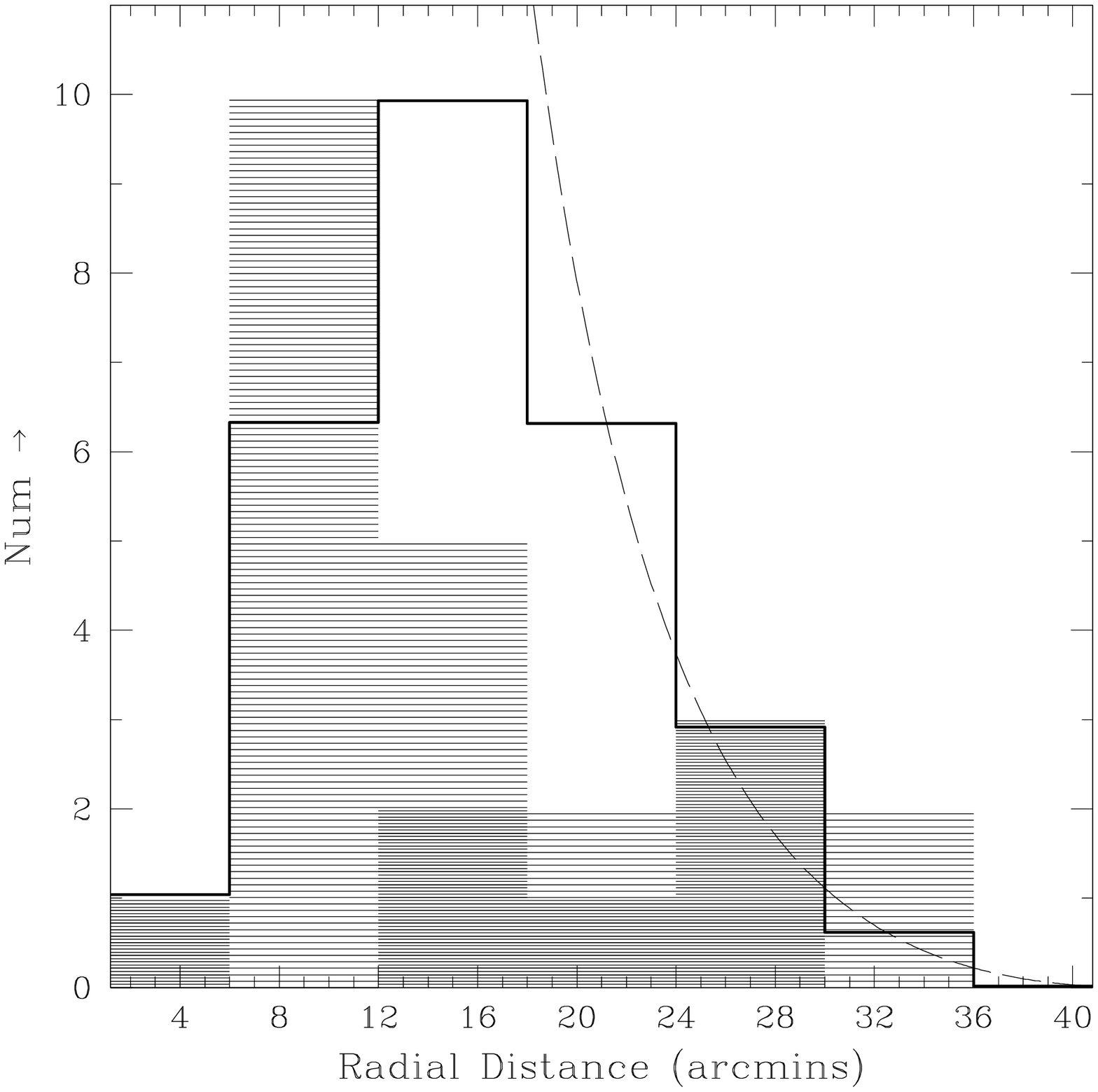}
\figcaption[Weldrake.fig19.eps]{Radial distribution of detected EcB stars. The total star distribution (open histogram) is overplotted to the same binning and normalised for comparison, along with a dotted line indicating the theoretical King Profile using parameters determined from \citet{Harris96}. The contact binaries (lighter histogram) are clearly distributed closer to the core than the main stellar population, with the detached EcB systems (dark histogram) being more segregated to the outer regions of the cluster.\label{EcBdist}}

\plotone{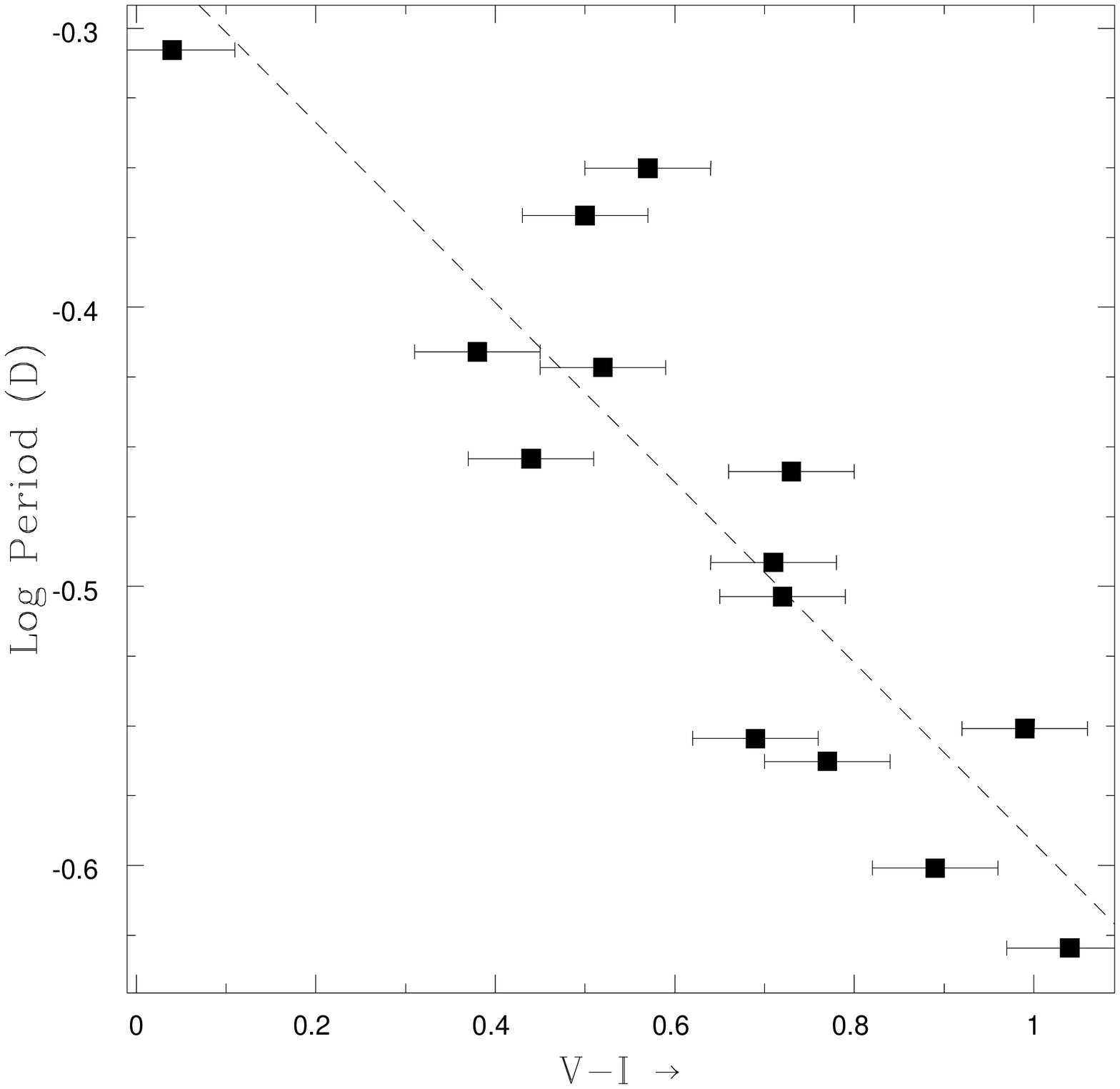}
\figcaption[Weldrake.fig20.eps]{Period-Colour diagram of our detected contact binaries. Binary systems are redder the shorter their period. A least-squares fit has been overplotted for completeness.\label{EcBPcol}}

\plotone{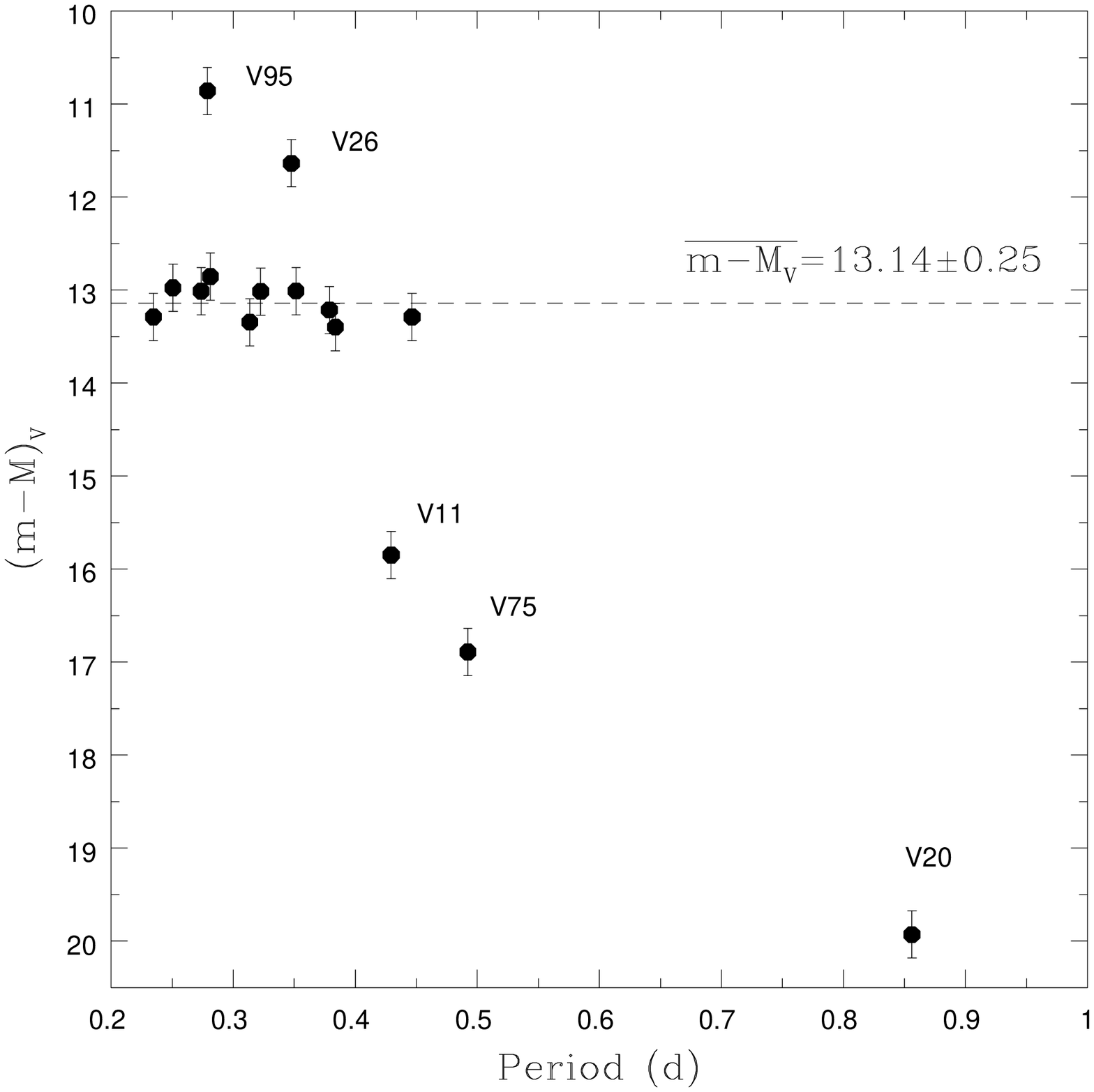}
\figcaption[Weldrake.fig21.eps]{The period versus distance modulus diagram for our sample of contact eclipsing binaries. Only those stars with complete colour information and a period $\leq$1 day are plotted. The average distance modulus of these 47 Tuc contact binaries is shown as the dashed line. Those systems that are not apparent members of the cluster are identified. V95 and V26 appear to be foreground systems in the Galactic Halo, whereas V11, V75 and V20 are more likely distant SMC members.The errorbars are the errors associated with the M$_V$ calculation, and include errors in the colour determination.\label{Ruccal}}

\plotone{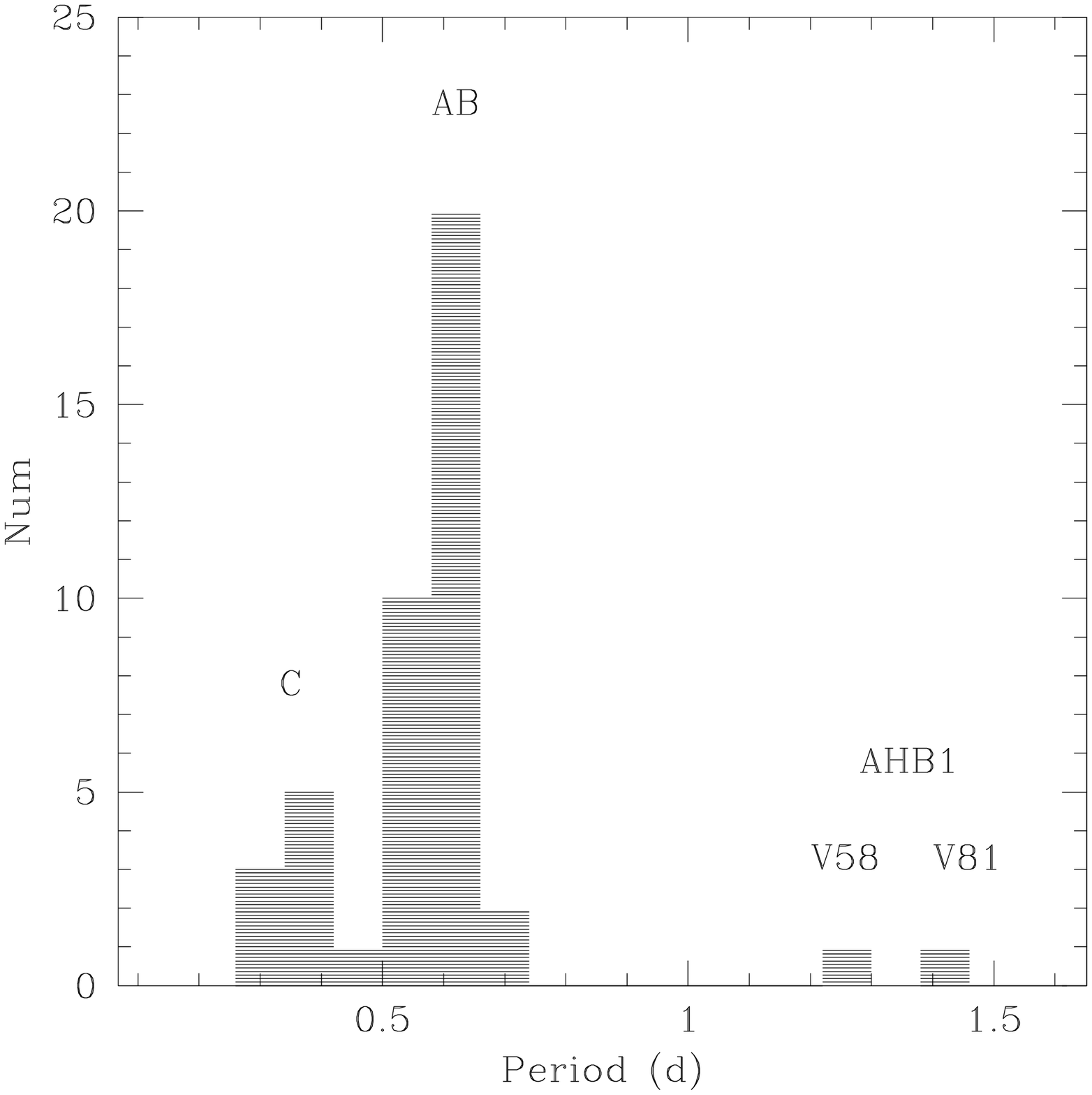}
\figcaption[Weldrake.fig22.eps]{Period distribution of detected RR Lyrae stars. Three populations are apparent, and marked accordingly. Two examples of AHB1 stars (both Halo stars) are marked and identified.\label{RRLyrP}}

\plotone{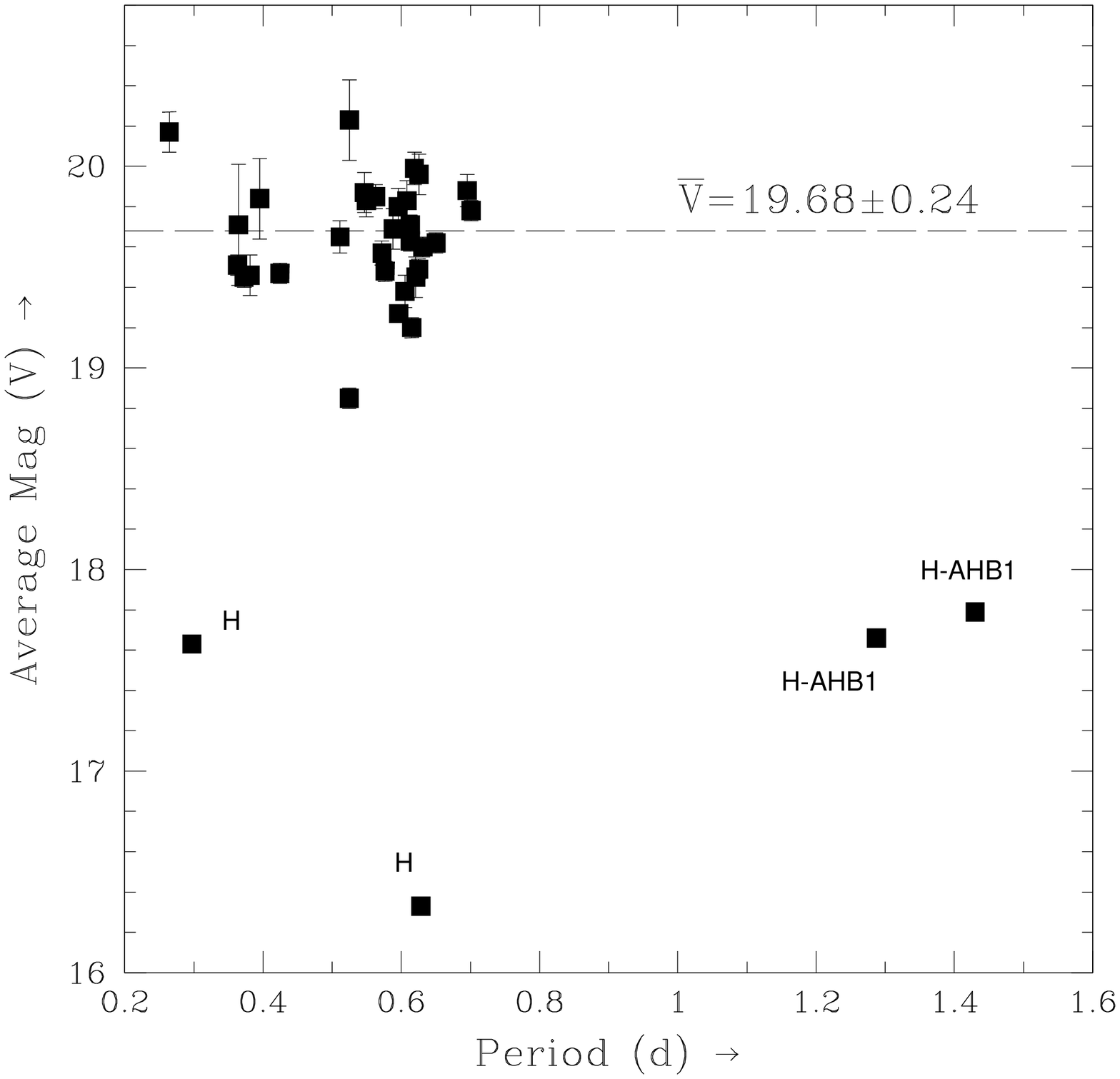}
\figcaption[Weldrake.fig23.eps]{Period-Luminosity diagram of those RR Lyraes for which we have magnitude information. The average V of our SMC RR Lyraes is indicated with a dashed line, along with the identification of Halo stars (H). AHB1 denotes the long-period RR Lyraes described in the text.\label{RRLyrPV}}

\plotone{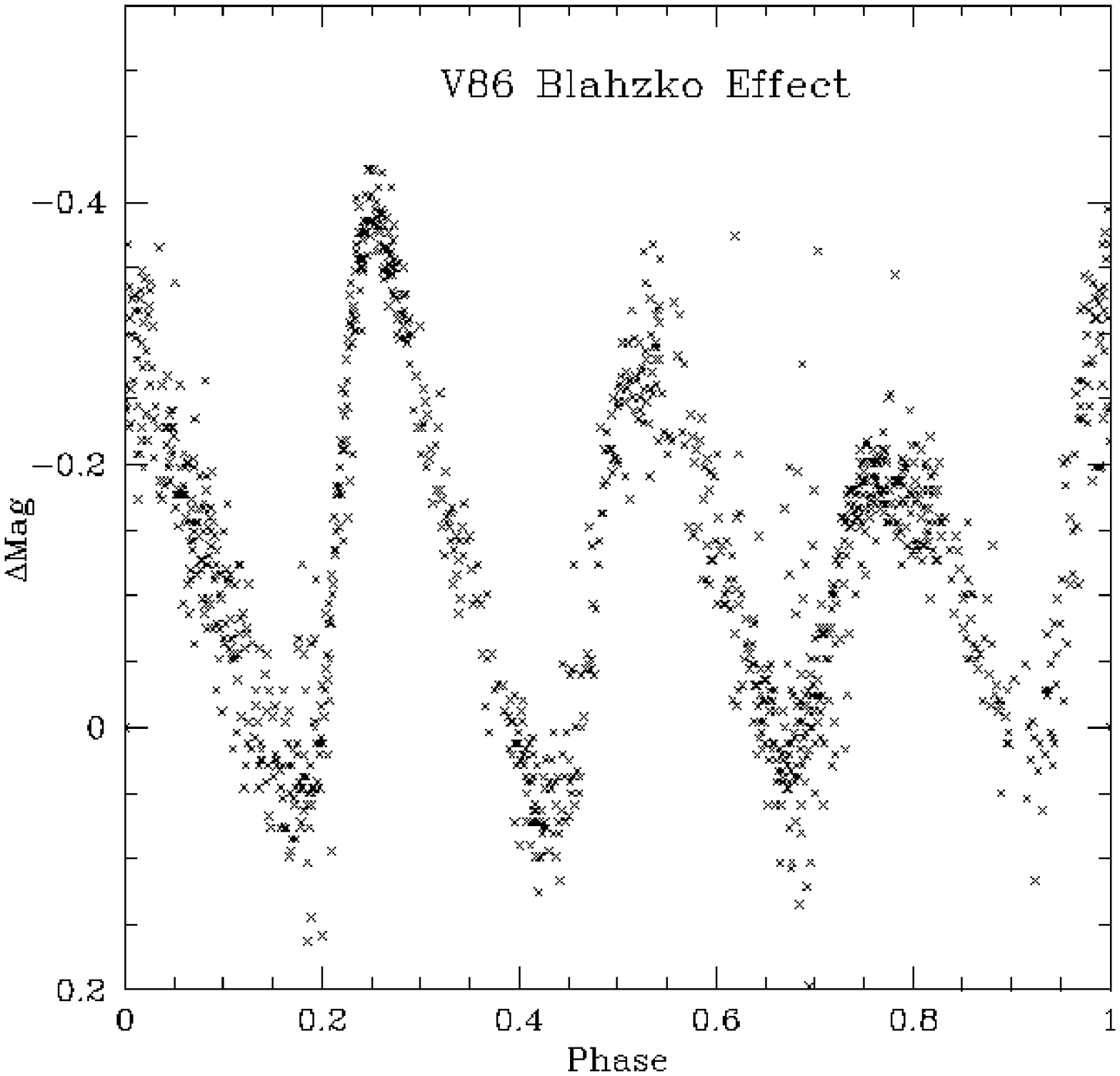}
\figcaption[Weldrake.fig24.eps]{Phase-wrapped lightcurve of SMC RR Lyrae V86, showing 4 periods. This star displays a remarkably symmetric Blahzko Effect, every 3 periods the amplitude of variation is significantly lower.\label{V86B}}

\clearpage

\begin{center}
\footnotesize
\begin{tabular}{lll} \hline
\noalign{\medskip}
$CCD$ & $\it{RA}(J2000.0)$ & $\it{DEC}(J2000.0)$ \\
& h m s & $^{\circ}$ $'$ $''$ \\
\noalign{\medskip}
\hline
\noalign{\medskip}
1 & 00:26:40 & $-$71:44:14 \\
2 & 00:26:40 & $-$71:57:42 \\
3 & 00:26:40 & $-$72:10:46 \\
4 & 00:26:40 & $-$72:23:42 \\
5 & 00:21:02 & $-$72:24:20 \\
6 & 00:21:02 & $-$72:11:12 \\
7 & 00:21:02 & $-$71:58:20 \\
8 & 00:21:02 & $-$71:45:15 \\
\noalign{\medskip}
\hline\hline
\end{tabular}  
\end{center}
\vspace*{3mm}                            
\normalsize{Table 1: Equatorial coordinates (J2000.0) for the centres of our CCDs.}

\footnotesize
\begin{tabular}{llllll} \hline\hline
\noalign{\medskip}
$ID$ & $Type$ & $RA(J2000.0)$ & $DEC(J2000.0)$ & $OGLEGC$ & $CCD$ \\
&& h m s & $^{\circ}$ $'$ $''$ &&\\
\noalign{\medskip}
\hline
\noalign{\medskip}
V1 & RR Lyr & 00:28:52.205 & $-$72:09:31.00 & - & 3 \\
V2 & RR Lyr & 00:28:56.555 & $-$72:08:18.75 & - & 3 \\
V3 & RR Lyr & 00:28:15.293 & $-$72 06:11.23 & - & 3 \\
V4 & LPV & 00:26:19.372 & $-$72:14:02.10 & OGLEGC254 & 3 \\
V5 & RR Lyr & 00:26:09.379 & $-$72:12:28.64 & OGLEGC255 & 3 \\
V6 & EcB & 00:26:10.514 & $-$72:11:07.75 & OGLEGC227 & 3 \\
V7 & EcB & 00:26:08.836 & $-$72:07:01.45 & OGLEGC228 & 3 \\
V8 & LPV & 00:25:38.176 & $-$72:16:37.62 & OGLEGC252 & 3 \\
V9 & LPV & 00:25:34.105 & $-$72:09:58.17 & OGLEGC222 & 3 \\
V10 & RR Lyr & 00:25:42.014 & $-$72:06:01.43 & OGLEGC223 & 3 \\
V11 & EcB & 00:27:33.659 & $-$72:04:38.26 & - & 3 \\
V12 & EcB & 00:26:42.960 & $-$72:15:20.71 & OGLEGC253 & 3 \\
V13 & RR Lyr & 00:26:56.834 & $-$72:10:13.19 & OGLEGC232 & 3 \\
V14 & EcB & 00:25:11.146 & $-$72:12:14.42 & OGLEGC250 & 3 \\
V15 & LPV & 00:25:06.187 & $-$72:10:23.48 & OGLEGC219 & 3 \\
V16 & LPV & 00:24:31.778 & $-$72:09:29.45 & OGLEGC251 & 3 \\
V17 & EcB & 00:24:25.499 & $-$72:08:31.60 & - & 3 \\
V18 & RR Lyr & 00:29:20.024 & $-$72:00:07.21 & - & 2 \\
V19 & RR Lyr & 00:29:02.392 & $-$71:58:57.71 & - & 2 \\
V20 & EcB & 00:28:55.205 & $-$71:59:56.93 & - & 2 \\
V21 & EcB & 00:29:09.429 & $-$71:51:26.42 & - & 2 \\
V22 & RR Lyr & 00:28:31.478 & $-$72:02:07.5 & - & 2 \\
V23 & RR Lyr & 00:28:27.374 & $-$71:58:32.19 & - & 2 \\
V24 & TyII Ceph & 00:28:15.314 & $-$71:58:21.58 & - & 2 \\
V25 & LPV & 00:27:56.580 & $-$71:57:26.71 & - & 2 \\
V26 & EcB & 00:26:59.860 & $-$71:55:09.77 & - & 2 \\
V27 & LPV & 00:26:12.380 & $-$72:02:13.24 & OGLEGC230 & 2 \\
V28 & LPV & 00:26:02.920 & $-$72:03:02.94 & OGLEGC229 & 2 \\
V29 & LPV & 00:25:37.513 & $-$71:56:03.71 & - & 2 \\
V30 & EcB & 00:25:15.920 & $-$71:56:06.76 & OGLEGC238 & 2 \\
V31 & LPV & 00:24:42.540 & $-$71:59:23.92 & OGLEGC220 & 2 \\
V32 & EcB & 00:25:00.451 & $-$72:00:02.83 & OGLEGC221 & 2 \\
V33 & LPV & 00:24:36.217 & $-$71:57:19.26 & - & 2 \\
V34 & EcB & 00:24:49.296 & $-$71:56:19.84 & - & 2 \\
\noalign{\medskip}
\hline\hline
\end{tabular}  

\footnotesize
\begin{tabular}{llllll} \hline\hline
\noalign{\medskip}
$ID$ & $Type$ & $RA(J2000.0)$ & $DEC(J2000.0)$ & $OGLEGC$ & $CCD$ \\
&& h m s & $^{\circ}$ $'$ $''$ &&\\
\noalign{\medskip}
\hline
\noalign{\medskip}
V35 & $\delta$ Scuti & 00:29:06.307 & $-$72:28:48.03 & - & 4 \\
V36 & LPV & 00:28:55.595 & $-$72:19:50.77 & - & 4 \\
V37 & Cepheid & 00:28:24.641 & $-$72:28:37.04 & - & 4 \\
V38 & RR Lyr & 00:28:40.161 & $-$72:24:03.96 & - & 4 \\
V39 & Det.EcB & 00:28:21.873 & $-$72:21:28.41 & - & 4 \\
V40 & RR Lyr & 00:27:56.936 & $-$72:21:56.20 & - & 4 \\
V41 & Det.EcB & 00:27:45.610 & $-$72:23:00.27 & - & 4 \\
V42 & LPV & 00:27:38.760 & $-$72:21:46.11 & - & 4 \\
V43 & EcB & 00:27:58.730 & $-$72:18:47.36 & - & 4 \\
V44 & RR Lyr & 00:27:24.760 & $-$72:23:38.47 & - & 4 \\
V45 & RR Lyr & 00:26:56.533 & $-$72:22:08.39 & - & 4 \\
V46 & LPV & 00:26:55.296 & $-$72:21:31.04 & - & 4 \\
V47 & RR Lyr & 00:26:07.430 & $-$72:24:39.51 & - & 4 \\
V48 & RR Lyr & 00:24:48.950 & $-$72:26:42.95 & - & 4 \\
V49 & RR Lyr & 00:24:45.564 & $-$72:24:33.29 & - & 4 \\
V50 & LPV & 00:25:05.083 & $-$72:26:58.64 & - & 4 \\
V51 & EcB & 00:24:58.186 & $-$72:22:11.08 & OGLEGC249 & 4 \\
V52 & RR Lyr & 00:24:31.318 & $-$72:28:13.25 & - & 4 \\
V53 & LPV & 00:24:11.112 & $-$72:27:15.40 & - & 4 \\
V54 & $\delta$ Scuti & 00:24:01.445 & $-$72:26:57.49 & - & 4 \\
V55 & LPV & 00:23:28.058 & $-$72:26:36.56 & OGLEGC248 & 5 \\
V56 & EcB & 00:23:15.809 & $-$72:18:53.93 & OGLEGC244 & 5 \\
V57 & RR Lyr & 00:21:48.627 & $-$72:25:21.48 & - & 5 \\
V58 & RR Lyr & 00:21:53.866 & $-$72:22:22.83 & - & 5 \\
V59 & EcB & 00:19:26.139 & $-$72:28:52.87 & - & 5 \\
V60 & AGB Puls & 00:18:21.268 & $-$72:26:24.44 & - & 5 \\
V61 & EcB & 00:22:00.581 & $-$72:02:04.03 & OGLEGC214 & 7 \\
V62 & LPV & 00:22:11.470 & $-$71:59:17.88 & - & 7 \\
V63 & LPV & 00:21:11.387 & $-$72:00:57.68 & - & 7 \\
V64 & RR Lyr & 00:21:06.552 & $-$71:56:37.49 & - & 7 \\
V65 & RR Lyr & 00:20:49.302 & $-$72:03:09.62 & OGLEGC213 & 7 \\
V66 & RR Lyr & 00:20:17.035 & $-$71:57:26.36 & - & 7 \\
V67 & $\delta$ Scuti & 00:19:09.879 & $-$72:04:27.80 & - & 7 \\
\noalign{\medskip}
\hline\hline
\end{tabular}  

\begin{center}
\footnotesize
\begin{tabular}{llllll} \hline\hline
\noalign{\medskip}
$ID$ & $Type$ & $RA(J2000.0)$ & $DEC(J2000.0)$ & $OGLEGC$ & $CCD$ \\
&& h m s & $^{\circ}$ $'$ $''$ &&\\
\noalign{\medskip}
\hline
\noalign{\medskip}
V68 & RR Lyr & 00:18:39.969 & $-$72:03:58.19 & - & 7 \\
V69 & Det.EcB & 00:22:52.951 & $-$72:03:40.68 & - & 7 \\
V70 & RR Lyr & 00:22:32.785 & $-$71:59:30.94 & - & 7 \\
V71 & RR Lyr & 00:22:43.840 & $-$71:57:20.66 & OGLEGC234 & 7 \\
V72 & RR Lyr & 00:23:36.984 & $-$71:45:14.81 & - & 8 \\
V73 & RR Lyr & 00:21:30.496 & $-$71:46:31.43 & - & 8 \\
V74 & RR Lyr & 00:20:55.545 & $-$71:41:30.92 & - & 8 \\
V75 & EcB & 00:29:05.153 & $-$71:43:25.90 & - & 1 \\
V76 & RR Lyr & 00:28:50.459 & $-$71:43:27.06 & - & 1 \\
V77 & RR Lyr & 00:28:15.717 & $-$71:46:11.79 & - & 1 \\
V78 & Det.EcB & 00:28:21.065 & $-$71:43:40.72 & - & 1 \\
V79 & RR Lyr & 00:28:21.180 & $-$71:40:31.50 & - & 1 \\
V80 & $\delta$ Scuti & 00:26:42.279 & $-$71:44:07.98 & - & 1 \\
V81 & RR Lyr & 00:26:42.083 & $-$71:40:24.53 & - & 1 \\
V82 & RR Lyr & 00:25:58.078 & $-$71:46:57.09 & - & 1 \\
V83 & RR Lyr & 00:28:22.977 & $-$71:28:55.44 & - & 1 \\
V84 & RR Lyr & 00:25:11.427 & $-$71:43:35.59 & - & 1 \\
V85 & Det.EcB & 00:24:43.872 & $-$71:47:38.26 & OGLEGC240 & 1 \\
V86 & RR Lyr & 00:24:58.198 & $-$71:42:13.35 & - & 1 \\
V87 & RR Lyr & 00:28:23.004 & $-$71:28:22.70 & - & 1 \\
V88 & RR Lyr & 00:18:49.569 & $-$72:09:38.80 & - & 6 \\
V89 & Det.EcB & 00:18:54.254 & $-$72:06:32.35 & - & 6 \\
V90 & RR Lyr & 00:20:57.408 & $-$72:12:04.35 & - & 6 \\
V91 & RR Lyr & 00:21:51.437 & $-$72:10:49.96 & - & 6 \\
V92 & RR Lyr & 00:21:23.958 & $-$72:15:33.79 & OGLEGC243 & 6 \\
V93 & Det.EcB & 00:21:07.603 & $-$72:08:55.80 & - & 6 \\
V94 & RR Lyr & 00:22:33.373 & $-$72:13:49.70 & OGLEGC246 & 6 \\
V95 & EcB & 00:22:47.469 & $-$72:13:17.20 & OGLEGC245 & 6 \\
V96 & LPV & 00:22:40.876 & $-$72:09:22.51 & OGLEGC216 & 6 \\
V97 & EcB & 00:23:06.015 & $-$72:09:30.52 & - & 6 \\
V98 & RR Lyr & 00:23:25.283 & $-$72:19:35.85 & OGLEGC247 & 5 \\
V99 & RR Lyr & 00:25:49.582 & $-$71:58:27.35 & OGLEGC226 & 2 \\
V100 & EcB & 00:25:32.213 & $-$72:01:51.37 & OGLEGC225 & 2 \\ 
\noalign{\medskip}
\hline\hline
\end{tabular}
\end{center}
  
\vspace*{3mm}                            
\normalsize{Table 2: Table of all detected variable stars in the field of 47 Tuc. Equatorial coordinates given in J2000.0.  If the star is previously known, the OGLEGC number is given (Kaluzny et al, 1998). Those marked with a dash are therefore new discoveries. The type of the variable is noted for completion; along with the number of the CCD chip on which it was found. Thirteen of the variable stars found in CCD3 were also independently identified by Eduard Westra, and are presented in his masters thesis 'A search for planets and other variables in 47 Tucanae' \citep{Westra03}.}

\pagebreak

\begin{center}
\footnotesize
\begin{tabular}{llllll} \hline\hline
\noalign{\medskip}
$ID$ & $Location$ & $Period (d)$ & $V_{CMD}$ & $V-I_{CMD}$ & $Total Amp(mag_{V+R})$ \\
\noalign{\medskip}
\hline
\noalign{\medskip}
V6 & BS & 0.3788 & 16.73 & 0.52 & 0.45 \\
V7 & BS & 1.1506 & 16.25 & 0.34 & 0.4 \\
V11 & BMS & 0.4294 & 18.96 & 0.50 & 0.2 \\
V12 & BS & 0.4465 & 16.66 & 0.57 & 0.45 \\
V14 & BS & 0.3514 & 16.47 & 0.44 & 0.28 \\
V17 & - & 0.3005 & - & - & 0.25 \\
V20 & SMC.BS? & 0.8561 & 20.30 & 0.13 & 0.7 \\
V21 & BMS & 0.2812 & 18.88 & 0.99 & 0.75 \\
V26 & FGRND? & 0.3476 & 15.99 & 0.73 & 0.09 \\
V30 & BMS & 0.2506 & 18.56 & 0.89 & 0.15 \\
V32 & BMS & 0.3136 & 18.14 & 0.72 & 0.6 \\
V34 & - & 0.2155 & - & - & 0.2 \\
V39 & BMS & 4.6015 & 18.17 & 1.06 & 0.18 \\
V41 & - & 5.3648 & $\sim$21 & - & 1.0 \\
V43 & - & 0.2602 & - & - & $\sim$1 \\
V51 & MSTO & 0.3226 & 17.43 & 0.71 & 0.35 \\
V56 & BS & 0.3837 & 16.27 & 0.38 & 0.25 \\
V59 & - & 0.4618 & $\sim$21 & - & $\sim$1 \\
V61 & BMS & 0.2737 & 18.02 & 0.77 & 0.4 \\
V69 & MSTO & 5.2239 & 17.08 & 0.58 & 0.65 \\
V75 & FGRND? & 0.4922 & 18.04 & 0.04 & 0.18 \\
V78 & BMS & $?$ & 19.31 & 0.88 & 0.2 \\
V85 & - & 2.1559 & $\sim$20 & - & 0.55 \\
V89 & - & 1.6215 & $\sim$21 & - & 0.8 \\
V93 & BMS & $?$ & 19.71 & 0.93 & 0.2 \\
V95 & FGRND? & 0.2789 & 15.49 & 0.69 & 0.4 \\
V97 & - & 0.3973 & $\sim$19 & - & 0.3 \\
V100 & BMS & 0.2347 & 19.47 & 1.04 & 0.3 \\      
\noalign{\medskip}
\hline\hline
\end{tabular}
\end{center}
  
\vspace*{3mm}                            
\normalsize{Table 3: Table of EcB found in our sample. The likely type is noted as given by the location on the CMD, if data are available for that star. BS=Blue Straggler, BMS=Binary Main Sequence, FGRND=foreground, MSTO=Main Sequence Turnoff. The period is given in days, along with the V magnitude, V-I on the CMD data, and the total amplitude of the variation, measured in magnitudes with a combined Cousins V+R filter.}

\begin{center}
\footnotesize
\begin{tabular}{lllll} \hline\hline
\noalign{\medskip}
$ID$ & $Period (d)$ & $V_{CMD}$ & $V-I_{CMD}$ & $Total Amp(mag_{V+R})$ \\
\noalign{\medskip}
\hline
\noalign{\medskip}
V1 & 0.5052 & - & - & 0.8 \\
V2 & 0.3812 & 19.36 & 0.39 & 0.6 \\
V3 & 0.3647 & 19.36 & 0.35 & 0.7 \\
V5 & 0.5251 & 19.80 & 0.50 & 0.9 \\
V10 & 0.2971 & 17.60 & 0.33 & 0.25 \\
V13 & 0.3633 & 19.66 & 0.49 & 0.6 \\
V18 & 0.5794 & - & - & 0.6 \\
V19 & 0.5766 & 19.63 & 0.53 & 0.9 \\
V22 & 0.4244 & 19.52 & 0.37 & 0.6 \\
V23 & 0.7008 & 19.68 & 0.66 & 0.9 \\
V38 & 0.5495 & 19.60 & 0.17 & 0.9 \\
V40 & 0.5627 & 19.82 & 0.37 & 0.6 \\
V44 & 0.6190 & 19.91 & 0.64 & 0.8 \\
V45 & 0.6524 & - & - & 0.6 \\
V47 & 0.6129 & 19.88 & 0.63 & 0.7 \\
V48 & 0.6210 & 19.58 & 0.37 & 0.6 \\
V49 & 0.3727 & 19.60 & -0.13 & 0.4 \\
V52 & 0.5820 & 19.64 & 0.46 & 0.6 \\
V57 & 0.6084 & 19.66 & 0.58 & 0.7 \\
V58 & 1.2875 & 17.49 & 0.53 & 0.8 \\
V64 & 0.5963 & 19.30 & 0.84 & 0.35 \\
V65 & 0.6327 & - & - & 0.5 \\
V66 & 0.6503 & 19.74 & 0.73 & 0.4 \\
V68 & 0.6131 & 19.73 & 0.67 & 0.55 \\
V70 & 0.2653 & - & - & 0.6 \\
V71 & 0.6151 & 19.30 & 0.36 & 0.5 \\
V72 & 0.6227 & - & - & 0.55 \\
V73 & 0.6283 & 16.33 & 0.41 & 0.04 \\
V74 & 0.6291 & 19.73 & 0.49 & 0.7 \\
V76 & 0.5954 & 19.68 & 0.42 & 0.8 \\
V77 & 0.2645 & 20.00 & 0.49 & 0.7 \\
V79 & 0.6053 & 19.59 & 0.55 & 0.8 \\
V81 & 1.4301 & 17.82 & 0.49 & 0.75 \\
V82 & 0.6252 & 19.33 & 0.47 & 0.6 \\
\noalign{\medskip}
\hline\hline
\end{tabular}
\end{center}
  
\begin{center}
\footnotesize
\begin{tabular}{lllll} \hline\hline
\noalign{\medskip}
$ID$ & $Period (d)$ & $V_{CMD}$ & $V-I_{CMD}$ & $Total Amp(mag_{V+R})$ \\
\noalign{\medskip}
\hline
\noalign{\medskip}
V83 & 0.5465 & 19.97 & 0.51 & 1.0 \\
V84 & 0.5882 & 19.69 & 0.52 & 1.0 \\
V86 & 0.5247 & 19.05 & 0.54 & 0.5 \\
V87 & 0.3951 & 19.51 & 0.38 & 0.7 \\
V88 & 0.6313 & 19.52 & 0.46 & 0.7 \\
V90 & 0.6952 & 19.70 & 0.70 & 0.9 \\
V91 & 0.6533 & - & - & 0.3 \\
V92 & 0.6256 & 20.10 & 0.59 & 0.7 \\
V94 & 0.5721 & 19.65 & 0.80 & 0.9 \\
V98 & 0.5114 & 19.90 & 0.51 & 0.9 \\
V99 & 0.6475 & - & - & 0.25 \\  
\noalign{\medskip}
\hline\hline
\end{tabular}
\end{center}
  
\vspace*{3mm}                            
\normalsize{Table 4: Table of RR Lyraes found in our data. The ID, period, colour, V magnitude as per the CMD dataset, if known, and total variability amplitude are noted.  }

\begin{center}
\footnotesize
\begin{tabular}{lllll} \hline\hline
\noalign{\medskip}
$ID$ & $Period (d)$ & $V_{CMD}$ & $V-I{CMD}$ & $Total Amp(mag_{V+R}$ \\
\noalign{\medskip}
\hline
\noalign{\medskip}
V4 & $>$40 & 16.51 & 1.75 & 0.2 \\
V8 & $>$50 & 17.32 & 1.90 & $>$0.3 \\
V9 & $>$20 & 16.13 & 0.92 & $>$0.1 \\
V15 & $>$30 & 15.36 & 1.00 & 0.19 \\
V16 & 3.4629 & 16.72 & 1.05 & 0.18 \\
V24 & 0.3871 & 18.13 & 1.82 & 0.2 \\
V25 & 5.8524 & 16.72 & 0.97 & 0.1 \\
V27 & 2.416 & 17.57 & 1.12 & 0.08 \\ 
V28 & 4.159 & 15.00 & 1.03 & 0.13 \\ 
V29 & 4.639 & 18.79 & 1.64 & 0.1 \\
V31 & 10.21 & 16.23 & 0.90 & 0.18 \\
V33 & $>$50 & 16.57 & 1.51 & $>$0.2 \\ 
V35 & 0.0932 & $\sim$22 & - & 1.0 \\
V36 & 10.7 & 19.66 & 0.91 & 0.3 \\
V37 & 2.5724 & 16.97 & 0.71 & 0.9 \\
V42 & 18.3 & 16.66 & 1.49 & 0.04 \\
V46 & $>$40 & 16.39 & 1.89 & $>$0.15 \\
V50 & $>$100 & 16.90 & 2.34 & $>$0.2 \\
V53 & $>$100 & 16.57 & 1.73 & $>$0.4 \\
V54 & 0.0834 & $\sim$21 & - & 1.0 \\
V55 & $>$60 & 16.46 & 1.76 & $>$0.2 \\
V60 & 0.2544 & 17.20 & 2.99 & 0.07 \\
V62 & $>$50 & 17.56 & 2.03 & $>$0.45 \\
V63 & $>$30 & 17.09 & 1.46 & $>$0.08 \\
V67 & 0.0829 & $\sim$20 & - & 0.7 \\
V80 & 0.2144 & $\sim$22 & - & $\sim$1 \\
V96 & 8.235 & 16.56 & 1.14 & 0.07 \\
\noalign{\medskip}
\hline\hline
\end{tabular}
\end{center}
  
\vspace*{3mm}                            
\normalsize{Table 5: Table of LPV's and the other miscellaneous variables. The ID, period, colour, V magnitude as per the CMD dataset, if known, and total variability amplitude are noted.}

\end{document}